\documentclass[aps,twocolumn,prb,superscriptaddress,amsmath,amssymb]{revtex4-2} %% for PRL
\usepackage{dcolumn}
\usepackage{bm}
\usepackage{amsmath}
\usepackage{txfonts}
\usepackage[T1]{fontenc}
\usepackage{xspace}
\usepackage{ulem}
\usepackage{comment}
\newcommand{\bras}[1]{\langle#1\rvert}
\newcommand{\kets}[1]{\lvert#1\rangle}

\newcommand{\means}[1]{\langle#1\rangle}

\ifx\pdfoutput\undefined
\usepackage[dvipdfmx]{graphicx}
\usepackage[dvipdfmx]{hyperref}
\usepackage[dvipdfmx]{color}
\usepackage[dvipdfmx]{xcolor}
\else
\usepackage{graphicx}
\usepackage{hyperref}
\usepackage{color}
\usepackage{xcolor}
\fi
\begin{document}
\let\emph\textit

\title{
% Hall-type
Antisymmetric Thermopolarization by Electric Toroidicity
  }
\author{Joji Nasu}
\affiliation{
  Department of Physics, Tohoku University, Sendai 980-8578, Japan
}
\affiliation{
  PRESTO, Japan Science and Technology Agency, Honcho Kawaguchi, Saitama 332-0012, Japan
}
\author{Satoru Hayami}
\affiliation{
  Department of Applied Physics, The University of Tokyo, Tokyo 113-8656, Japan
}
\affiliation{
  PRESTO, Japan Science and Technology Agency, Honcho Kawaguchi, Saitama 332-0012, Japan
}

\date{\today}
\begin{abstract}
  We investigate electric polarizations emergent perpendicular to an applied thermal gradient in insulating systems.
  The thermally-induced electric polarization, known as thermopolarization, has been studied conventionally in the case that an electric polarization appears along the thermal gradient.
  Here, we focus on the antisymmetric component of the thermopolarization tensor and reveal that it becomes nonzero owing to the ferro-type order for electric-toroidal dipole moments.
  To describe local electric polarizations originating from the disproportionation of localized electronic clouds, we introduce a two-dimensional three-orbital model with localized $s$ and two $p$ orbitals, where the electric polarization at each site interacts with the neighboring one as dipole-dipole interactions.
  We find that a vortex-type configuration of local electric polarizations appears as a mean-field ground state, corresponding to a ferro-type electric-toroidal dipole order.
  By taking account of collective modes from this ordered state, we calculate the coefficient of the thermopolarization based on the linear response theory.
  The antisymmetric component is nonzero in the presence of the electric-toroidal dipole order.
  We clarify that fluctuations in the $p$ orbitals are crucial in enhancing the antisymmetric thermopolarization.
  We discuss the appearance conditions based on the symmetry argument and the relevance to real materials.
\end{abstract}
\maketitle

%%%%%%%%%%%
% Introduction

\section{Introduction}

The study of cross-correlations in condensed matter physics has a long history since discovering the magnetoelectric effect in Cr$_2$O$_3$~\cite{curie1894symetrie,Dzyaloshinski1960,astrov1960magnetoelectric,Folen1961}.
Even now, the coupling between quantities with distinct symmetries has attracted considerable attention in the fields of strongly correlated electron systems and multiferroics as it strongly reflects the nature of symmetry breaking~\cite{kimura2003magnetic,fiebig2005revival,Katsura2005,khomskii2006multiferroics,cheong2007multiferroics,khomskii2009trend}.
While the coupling between electricity and magnetism has been mainly studied, other properties such as elastic and thermal are also expected to contribute to cross-correlations.
For example, the thermal gradient trivially induces a thermal current in the system but can also generate an electric polarization and magnetization in systems with particular symmetries~\cite{wang2010thermally,Dyrdal2013,xiao2016thermoelectric,Dyrdal2018,Shitade2019quadrupole}.
The thermally-induced electric polarization is known as thermopolarization~\cite{Bresme2008,Wirnsberger2018,onishi2021pre}, where the difference of the temperatures at two opposite edges yields disproportionation of electronic clouds or lattice positions, and thereby, a macroscopic electric polarization appears in the system.
In this case, it is natural to consider that the direction of the electric polarization is parallel to that of the thermal gradient.
Nevertheless, one cannot exclude the possibility of the emergent polarization perpendicular to the thermal gradient, which is an anomalous contribution whose response tensor is not only off-diagonal but also antisymmetric similar to the Hall effect.

To elucidate in what cases the cross-correlation occurs, multipole-based research has developed~\cite{Hayami2018Classification,Suzuki2018,Watanabe2018group,Yatsushiro2021}.
It tells us the necessary conditions for the emergence of a cross-correlation response based on the symmetries of the lattice geometry, electronic structure, and order parameter.
Among them, toroidal-type orders have recently attracted increasing interest~\cite{spaldin2008toroidal,khomskii2009trend,Kopaev2009,Hayami2018Classification}.
In particular, a magnetic-toroidal dipole moment is crucial for magnetoelectric effects because it is odd for both time and spatial reversal operations.
The electric counterpart of the magnetic-toroidal multipoles can also be introduced, which is referred to as electric-toroidal multipoles~\cite{dubovik1990toroid,Johnson2012,Hlinka2016,cheong2018broken,Hayami2018Microscopic}.
The multipoles are given by the time-reversal even and axial tensors.
Recently, it was pointed out that the bond-length modulation emergent in the pyrochlore oxide Cd$_2$Re$_2$O$_7$~\cite{Hanawa2001,Jin2001,Hiroi2002,Yamaura2002,Castellan2002,Kendziora2005,Bari2003,Sergienko2003,Kobayashi2011,Yamaura2017,Hiroi2018,Matsubayashi2018} can be interpreted as an electric-toroidal quadrupole order, which is the spatial (time) reversal parity odd (even)~\cite{Matteo2017,Hayami2019ElectricToroidal}.
Nonetheless, an electronic order involving the electric-toroidal dipoles, simpler than the quadrupole ones, remains elusive.
This is because the dipole component is both spatial and time-reversal parity even, complicating its experimental observation while the longitudinal dissipationless spin-current generation was proposed recently, originating from electric-toroidal octupoles~\cite{hayami2021pre}.

On the other hand, an electric-toroidal dipole order caused by lattice distortions has been studied as a ferroaxial order~\cite{Hlinka2016}.
The ferroaxial (ferro-rotational) order was initially introduced as a ferro-type order described by an axial vector without the time and spatial symmetry breakings.
Recently, the attempt to observe the domains of the ferroaxial order has been made by light.
In the ferroaxial order, the mirror symmetry is preserved on the plane perpendicular to the ordering vector.
Once the electric field parallel to this vector breaks the mirror symmetry, a chirality appears in the system.
The optical rotation can identify the direction~\cite{hayashida2020visualization}.
This scheme directly observes the chirality induced by the electric field rather than the electric-toroidal dipole order.
Moreover, a ferroaxial order was also observed by using the second-harmonic generation via its electric quadrupole component~\cite{jin2020observation}.
Therefore, the direct observation of the electric-toroidal dipole is desired as a linear response.
However, this is not expected to couple linearly with electric and magnetic fields.

In this paper, we propose that the thermal response can be an appropriate probe to observe the electric-toroidal dipole originating from electronic orbitals.
We introduce a three-orbital model with localized $s$ and two $p$ orbitals capable of generating an electric polarization.
To consider the electric order of the local polarizations constituting a ferro-type electric-toroidal dipole configuration, we define the model Hamiltonian on a square-octagon lattice.
This is one of the simplest lattice structures to stabilize the electric-toroidal dipole order induced by the dipole-dipole interaction.
We examine the three-orbital model using the mean-field approximation and calculate the thermal response by applying excitation-wave theory.
The ferro-type electric-toroidal dipole order appears when the energy gap between the $s$ and $p$ orbitals is small compared with the energy scale of the dipole-dipole interaction.
We find that the macroscopic polarization appears perpendicular to the applied thermal gradient and its linear response coefficient is antisymmetric for their directions.
This effect is regarded as an antisymmetric thermopolarization, which is an intrinsic one unrelated to the relaxation time of the thermal transport.
We also clarify that the thermopolarization is strongly enhanced when the $p$ orbital level is lower than the $s$ orbital one.
This implies that fluctuations on the two $p$ orbitals play a crucial role in enhancing the thermopolarization.
We demonstrate the presence of the fluctuations by calculating the excitation spectrum, in which the low-energy excitations changing the direction of the local electric moment exist.
We also discuss the relevance to real materials and the origin of the antisymmetric thermopolarization based on the symmetry argument.

This paper is organized as follows.
In the next section, we introduce a three-orbital model on a two-dimensional lattice with local electric dipole moments.
The method used in this study is presented in Sec.~\ref{sec:method}.
In Sec.~\ref{sec:method_MF}, we describe the mean-field theory applied to the model Hamiltonian and the way to address the fluctuations from the mean fields as elementary excitations.
The formulation of the thermopolarization is given in Sec.~\ref{sec:form_therm}.
The results are shown in Sec.~\ref{sec:result}.
In Sec.~\ref{sec:result_electric_toroidal}, we present the mean-field results where the electric-toroidal dipole order appears when the orbital level splitting is small compared with the dipole-dipole interactions.
We also show that the antisymmetric thermopolarization emerges in the electric-toroidal dipole ordered phase in Sec.~\ref{sec:result_thermo}.
In Sec.~\ref{sec:result_excitations}, we show the elementary excitations, which are crucial for the emergence of the nonzero thermopolarization.
In Sec.~\ref{sec:discussion}, we discuss the relevance to real materials and the origin of the antisymmetric thermopolarization from the viewpoint of the symmetry.
Finally, Sec.~\ref{sec:summary} is devoted to the summary.

%%%%%%%%%%%
% Model & method

\section{Model}\label{sec:model}

\begin{figure}[t]
\begin{center}
\includegraphics[width=\columnwidth,clip]{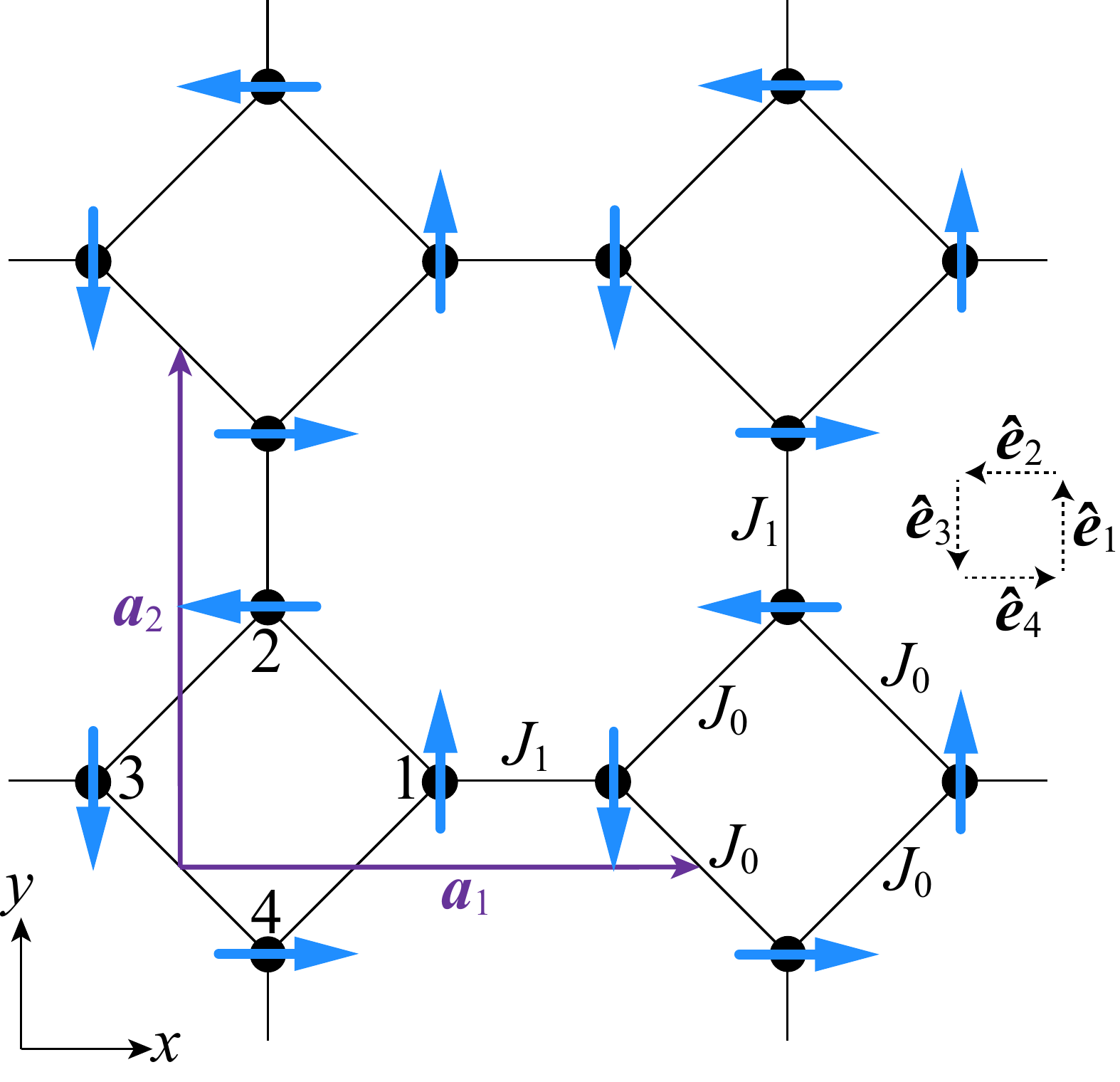}
\caption{
  Schematic figure of the two-dimensional square-octagon lattice.
  The vectors $\bm{a}_1$ and $\bm{a}_2$ with purple color represent primitive translational vectors, and there are four sites in a unit cell.
  The blue arrows stand for the local electric polarizations arranged with a toroidal-type configuration.
  The inset in the right side shows the four unit vectors, each of which is parallel to the blue arrow located on the corresponding site.
}
\label{fig:lattice}
\end{center}
\end{figure}

We introduce a simple model with local electric dipole moments to address the thermopolarization induced by the electric-toroidal dipole moment.
We consider three states, $s$, $p_x$, and $p_y$ orbitals, at each site on a two-dimensional square-octagon lattice, where the edges of neighboring squares are connected by lines as shown in Fig.~\ref{fig:lattice}.
Using these states, the local electric dipole moment on the $xy$ plane is given by $\bm{P}_i=(P_i^x,P_i^y)$ with $P_i^\gamma \propto \kets{s}_i\bras{p_\gamma}_i+{\rm H.c.}$ ($\gamma=x,y$).
Here, we assume that the constant of the proportionality is 1, namely, the elementary electric charge $e$ is regarded to be unity in addition to the reduced Planck constant $\hbar$, Boltzmann constant $k_B$, and the length of the primitive translational vectors.
The electric dipoles interact as the dipole-dipole interaction, which is given by
\begin{align}
  {\cal H}_{\rm int} = \sum_{\means{ij}}J_{ij}\left[\bm{P}_i\cdot\bm{P}_j -3(\bm{P}_i\cdot\bm{e}_{ij})(\bm{P}_j\cdot\bm{e}_{ij})\right],
  \label{eq:Hint}
\end{align}
where $\means{ij}$ stands for neighboring sites connected by the vector $\bm{e}_{ij}$ on the bonds of the square-octagon lattice, and $J_{ij}=J_0$ ($J_1$) for the intra-(inter-)square interaction, which is positive (see Fig.~\ref{fig:lattice}).
In addition to the dipole-dipole interaction, we consider the energy difference $\Delta$ between the $s$ and $p$ orbitals, and the local anisotropy, where the energy level of the $p$ orbital spread along the direction to the center of the square to which the site belongs is higher than the perpendicular orbital by $A (>0)$.
We refer to the former (latter) $p$ orbital as $\kets{p_{\perp}}$ ($\kets{p_{\parallel}}$).
These contributions are written as
\begin{align}
  {\cal H}_{\rm loc}= \sum_i \left[(\Delta +A)\kets{p_\perp}_i\bras{p_\perp}_i+\Delta\kets{p_\parallel}_i\bras{p_\parallel}_i\right].
\end{align}
The model Hamiltonian ${\cal H}={\cal H}_{\rm int}+{\cal H}_{\rm loc}$ is expected to exhibit the electric-toroidal dipole moment in a unit cell shown in Fig.~\ref{fig:lattice}.
If this electric-polarization configuration is present, the first (second) term in Eq.~\eqref{eq:Hint} disappears on the intra-(inter-)square bonds, and the ferro-(antiferro-)type contribution survives, which stabilize the assumed configuration in addition to the positive anisotropy $A$.
Thus, we believe that the present Hamiltonian provides a simple and appropriate model to discuss the effect of electric-toroidal dipole moments.

\section{Method}
\label{sec:method}

\subsection{Mean-Field Theory and Elementary Excitations}
\label{sec:method_MF}

To examine the electric-toroidal dipole order appearing in the Hamiltonian, we apply the mean-field approximation and linear excitation-wave theory.
The present system is similar to that with quantum paraelectricity, which has been discussed using the transverse Ising model~\cite{DEGENNES1963132,hemberger1996quantum,Prosandeev1999};
in this study, ${\cal H}_{\rm loc}$ is regarded as a transverse field because the local electric moment describes the mixing of the $s$ and $p$ orbitals.
One of the simplest ways to deal with the dynamics of the high-dimensional transverse Ising model is the linear excitation-wave approximation introduced later, and we apply this method to the present model.

The dipole-dipole interaction is symbolically written as
\begin{align}
  {\cal H}_{\rm int} = \sum_{\means{ij}}\sum_{\gamma\gamma'}J_{ij}^{\gamma\gamma'}P_i^{\gamma}P_j^{\gamma'}.
\end{align}
The mean-field approximation applied to it gives
\begin{align}
  {\cal H}_{\rm int}^{\rm MF} = \sum_{\means{ij}}\sum_{\gamma\gamma'}J_{ij}^{\gamma\gamma'}
  \left(
    \means{P_i^{\gamma}}P_j^{\gamma'} + P_i^{\gamma}\means{P_j^{\gamma'}}
     -\means{P_i^{\gamma}}\means{P_j^{\gamma'}}
  \right),
\end{align}
where the different mean-fields are prepared for four sublattice sites ($M=4$) in the unit cell of the square-octagon lattice (see Fig.~\ref{fig:lattice}).
We determine the mean-fields $\means{P_i^{\gamma}}$ by solving the single-site Hamiltonian ${\cal H}_i^{\rm MF}$ in ${\cal H}^{\rm MF} = {\cal H}_{\rm int}^{\rm MF} + {\cal H}_{\rm loc}$, and the expectation value is calculated for the ground state of ${\cal H}^{\rm MF}_i$, $\kets{0}_i$.

Next, we introduce the linear excitation-wave theory.
The deviation from the mean-field Hamiltonian is written as
\begin{align}
  {\cal H}'={\cal H}_{\rm int}-{\cal H}_{\rm int}^{\rm MF}=
  \sum_{\means{ij}}\sum_{\gamma\gamma'}J_{ij}^{\gamma\gamma'}\delta P_i^{\gamma} \delta P_j^{\gamma'},
\end{align}
where $\delta P_i^{\gamma}=P_i^{\gamma}-\means{P_i^{\gamma}}$.
In the linear excitation-wave theory, $\delta P_i^{\gamma}$ is approximated by extracting the matrix elements involving the ground state of ${\cal H}_i^{\rm MF}$ as~\cite{Onufrieva1985,Papanicolaou1988367,doi:10.1143/JPSJ.70.3076,doi:10.1143/JPSJ.72.1216,PhysRevB.60.6584,PhysRevB.88.224404,PhysRevB.88.205110}
\begin{align}
  \delta P_i^{\gamma}\simeq \sum_{m=1,2} a_{im}^\dagger \bras{m}_i\delta P_i^{\gamma}\kets{0}_i + {\rm H.c.}, 
\end{align}
where $\kets{m}_i$ with $m=1,2$ is the excited state of the local mean-field Hamiltonian ${\cal H}_i^{\rm MF}$ at site $i$, and $a_{mi}^\dagger=\kets{m}_i\bras{0}_i$ is assumed to be a creation operator of boson, corresponding to the Holstein-Primakoff quasiparticle.
Applying this approximation, we rewrite the Hamiltonian as a bilinear form of the bosonic operators:
\begin{align}
  {\cal H}\simeq \tilde{\cal H} = \frac{1}{2}\sum_{\bm{k}ll'}H_{{\bm{k}}ll'}{\cal A}_{\bm{k}l}^\dagger {\cal A}_{\bm{k}l'},
\end{align}
where ${\cal A}_{\bm{k}}^\dagger =(a_{\bm{k}1}^\dagger,\cdots, a_{\bm{k},2M}^\dagger, a_{-\bm{k}1},\cdots, a_{-\bm{k},2M})$ whose element is assigned by $l=1,2,\cdots 4M$, and $a_{\bm{k}(s,m)}^\dagger = \sqrt{M/N}\sum_{i\in s} a_{im}^\dagger e^{i\bm{k}\cdot {\bm{r}_i}}$ for sublattice $s$.
The $4M\times 4M$ matrix $H_{\bm{k}}$ is diagonalized by the Bogoliubov transformation with the paraunitary matrix $T_{\bm{k}}$ as~\cite{COLPA1978327}
\begin{align}
  \tilde{\cal H}=\frac{1}{2}\sum_{\bm{k}n} {\cal E}_{\bm{k}n}{\cal B}_{\bm{k}n}^\dagger {\cal B}_{\bm{k}n}+{\rm const.},
\end{align}
where ${\cal E}_{\bm{k}}=(\varepsilon_{\bm{k}1},\cdots, \varepsilon_{\bm{k},2M}, \varepsilon_{-\bm{k}1},\cdots, \varepsilon_{-\bm{k},2M})$ and ${\cal B}_{\bm{k}}^\dagger=(b_{\bm{k}1}^\dagger,\cdots, b_{\bm{k},2M}^\dagger, b_{-\bm{k}1},\cdots, b_{-\bm{k},2M})={\cal A}_{\bm{k}}^\dagger T_{\bm{k}}^{-1\dagger}$.
The paraunitary matrix satisfies the following equation 
\begin{align}
  T_{\bm{k}} {\cal I}T_{\bm{k}}^\dagger=T_{\bm{k}}^\dagger {\cal I}T_{\bm{k}}={\cal I},
\end{align}
where ${\cal I}$ is the paraunit matrix, which is diagonal and defined such that ${\cal I}_{nn}={\cal I}_{n}=+1$ for $n\le 2M$ and ${\cal I}_{n}=-1$ for $n>2M$.

\subsection{Formalism of Thermopolarization}
\label{sec:form_therm}

Here, we introduce the coefficient of the off-diagonal thermopolarization, $\beta^{xy}$, defined as
\begin{align}
  \frac{\means{P^x}_{\nabla_y T}}{V} = \beta^{xy} (-\nabla_y T),
\end{align}
where $\means{P^x}_{\nabla_y T}$ stands for the polarization under the thermal gradient, and $V$ is volume.
When the thermal gradient is absent, the averaged macroscopic polarization $\means{P^x}$ becomes zero, i.e, $\means{P^x}=0$ in equilibrium because of the toroidal-type configuration of the local electric polarizations shown in Fig.~\ref{fig:lattice}.
Since $P^x$ should vanish without bosonic excitations, the total polarization is approximately written as a bilinear form of the bosons:
\begin{align}
  P^x=\sum_i P_i^x\simeq \frac{1}{2}\sum_{\bm{k}nn'} {\cal P}_{\bm{k}nn'}^x {\cal B}_{\bm{k}n}^\dagger {\cal B}_{\bm{k}n'},
  \label{eq:Px}
\end{align}
where ${\cal P}_{\bm{k}}^x$ is a $4M\times 4M$ Hermitian matrix.
The velocity matrix is given by 
\begin{align}
  {\cal V}^y_{\bm{k}}=T_{\bm{k}}^\dagger \frac{\partial H_{\bm{k}}}{\partial k_y} T_{\bm{k}}.
  \label{eq:Vy}
\end{align}
Using these quantities, the coefficient $\beta^{xy}$ is represented as~\cite{Murakami_Okamoto2017,Shitade2019,Li_Mook2020}
\begin{align}
  \beta^{xy}=-\frac{1}{V}
\sum_{\bm{k}}\sum_{n=1}^{2M}c_1(n(\varepsilon_{\bm{k}n}))\Omega_{\bm{k}n}^{xy},
\label{eq:betaxy}
\end{align}
where the temperature-independent part $\Omega_{\bm{k}n}^{xy}$ is given by
\begin{align}
  \Omega_{\bm{k}n}^{xy}
  =-2\sum_{n'(\ne n)}^{4M}\frac{{\rm Im}[{\cal P}_{\bm{k}nn'}^{x}{\cal V}_{\bm{k}n'n}^{y}]{\cal I}_n {\cal I}_{n'}}{({\cal I}_n\varepsilon_{\bm{k}n}-{\cal I}_{n'}\varepsilon_{\bm{k}n'})^2}.
  \label{eq:Omegaxy}
\end{align}
The temperature dependence of $\beta^{xy}$ originates from its coefficient $c_1(n(\varepsilon_{\bm{k}n}))$, where $n(\varepsilon)=(e^{\varepsilon/T}-1)^{-1}$ is the Bose distribution function, and
\begin{align}
  c_1(x)=(1+x)\ln(1+x)-x\ln x.
\end{align}
We only consider the antisymmetric part, i.e., $\beta^{xy}=-\beta^{yx}$, which is an intrinsic contribution independent of the relaxation time.

\section{Result}
\label{sec:result}

\subsection{Electric Toroidal Dipole Order}
\label{sec:result_electric_toroidal}

\begin{figure}[t]
\begin{center}
\includegraphics[width=\columnwidth,clip]{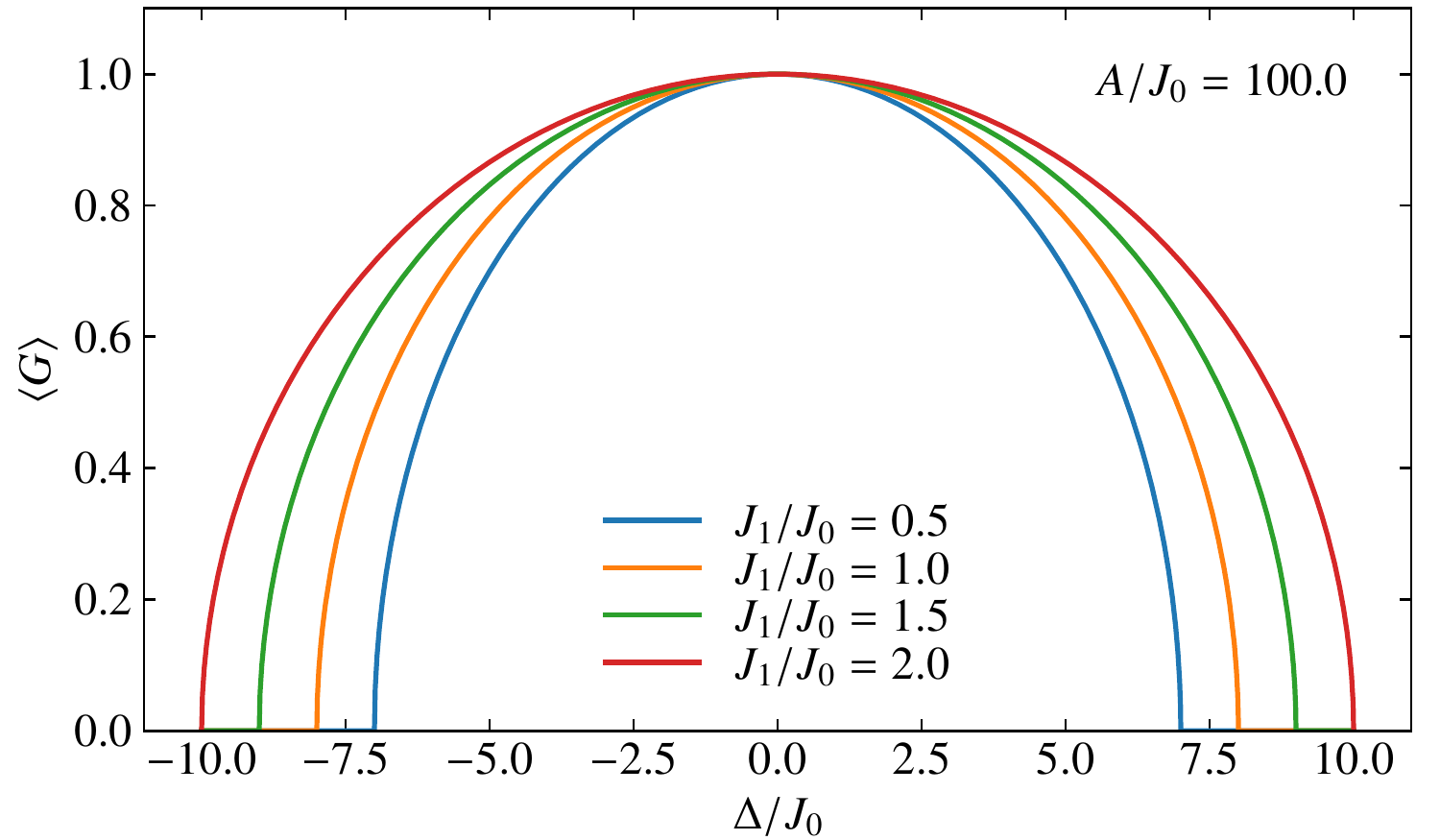}
\caption{
$\Delta$ dependence of the order parameter of the toroidal-type configuration for the electric dipole moments at $A/J_0=100$.
}
\label{fig:moment}
\end{center}
\end{figure}

First, we show the mean-field results at zero temperature.
In the present calculations, we only find the toroidal-type electric dipole order depicted in Fig.~\ref{fig:lattice} as expected.
The order parameter is given by
\begin{align}
  \means{G}=\frac{1}{N}\sum_{p=1,2,3,4}\hat{\bm{e}}_p\cdot\sum_{i\in p}\means{\bm{P}_i},
\end{align}
where we define the following unit vectors: $\hat{\bm{e}}_1=(0,1)$, $\hat{\bm{e}}_2=(-1,0)$, $\hat{\bm{e}}_3=(0,-1)$, and $\hat{\bm{e}}_4=(1,0)$ (see the inset of Fig.~\ref{fig:lattice}).
Figure~\ref{fig:moment} shows the $\Delta$ dependence of $\means{G}$ for several values of $J_1$ at $A/J_0=100$.
This quantity is nonzero around $\Delta=0$, indicating the electric-toroidal dipole order, and it continuously decreases and becomes zero by increasing $|\Delta|$. 
As discussed in Sec.~\ref{sec:method_MF}, the present model is similar to the transverse Ising model.
For the case of the large anisotropy, the local $s$ and $p_{\parallel}$ orbitals with the energy difference $\Delta$ dominate the low-energy properties, and nonzero $\means{G}$ is the consequence of the mixing between these orbitals.
The symmetric and dome-like behavior of $\means{G}$ as a function of $\Delta$ is understood as an analogy of the transverse Ising model.
The thermopolarization in this model on the zigzag chain is discussed in Appendix~\ref{app:TI}.

\begin{figure}[t]
\begin{center}
\includegraphics[width=\columnwidth,clip]{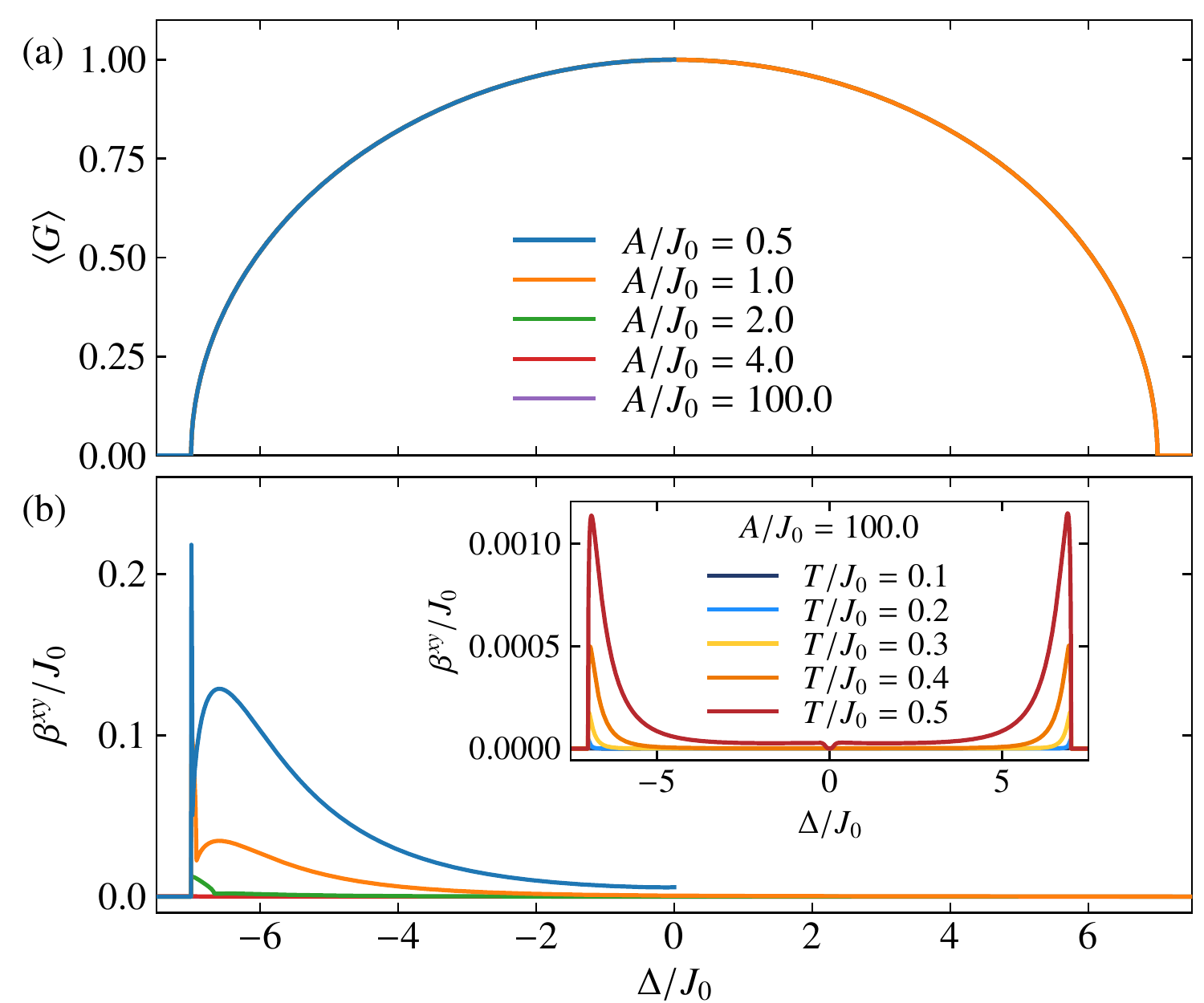}
\caption{
  (a) Toroidal order parameter and (b) the coefficient of the thermopolarization, $\beta^{xy}$, as functions of $\Delta$ for several $A$.
  In (b), the temperature is set at $T/J_0=0.3$.
The inset of (b) shows the $\Delta$ dependence of $\beta^{xy}$ for several temperatures at $A/J_0=100$.
$J_1/J_0$ is fixed to $0.5$.
We show the data only for the region where the ground state is stable against the creation of elementary excitations.
}
\label{fig:g_beta}
\end{center}
\end{figure}

As shown in Fig.~\ref{fig:moment}, the region of the electric-toroidal dipole order becomes large with increasing $J_1$.
The critical value of $\Delta$ is given by $\Delta_c=6J_0+2J_1$, which is understood from the magnitude of mean field yielded by the electric dipoles surrounding a certain site.
Note that $\Delta_c$ is independent of the local anisotropy $A$.
To confirm this clearly, we show the $A$ dependence of the order parameter $\means{G}$ in Fig.~\ref{fig:g_beta}(a).
In this figure, $\means{G}$ as a function of $\Delta$ is presented for several values of $A$, but all the lines appear to overlap with each other.
This result indicates that the anisotropy does not affect not only the phase boundary but also the $\Delta$ dependence of $\means{G}$ even for small $A$.

\subsection{Thermopolarization}
\label{sec:result_thermo}

Although the ground-state phase diagram remains largely intact for the anisotropy, it is expected to change the excitation spectra.
The low-energy excitations from the ground state contribute to the transport phenomena.
In particular, we focus on the off-diagonal thermopolarization, which was introduced in the previous section.
Figure~\ref{fig:g_beta}(b) shows the $\Delta$ dependence of the coefficient $\beta^{xy}$ for several $A$ at $T/J_0 =0.3$. 
As shown in this figure, $\beta^{xy}$ is almost zero at $A/J_0=100$, but it increases with decreasing $A$.
We find that $\beta^{xy}$ takes a large value near the critical point in the region of the negative $\Delta$.

\begin{figure}[t]
\begin{center}
\includegraphics[width=\columnwidth,clip]{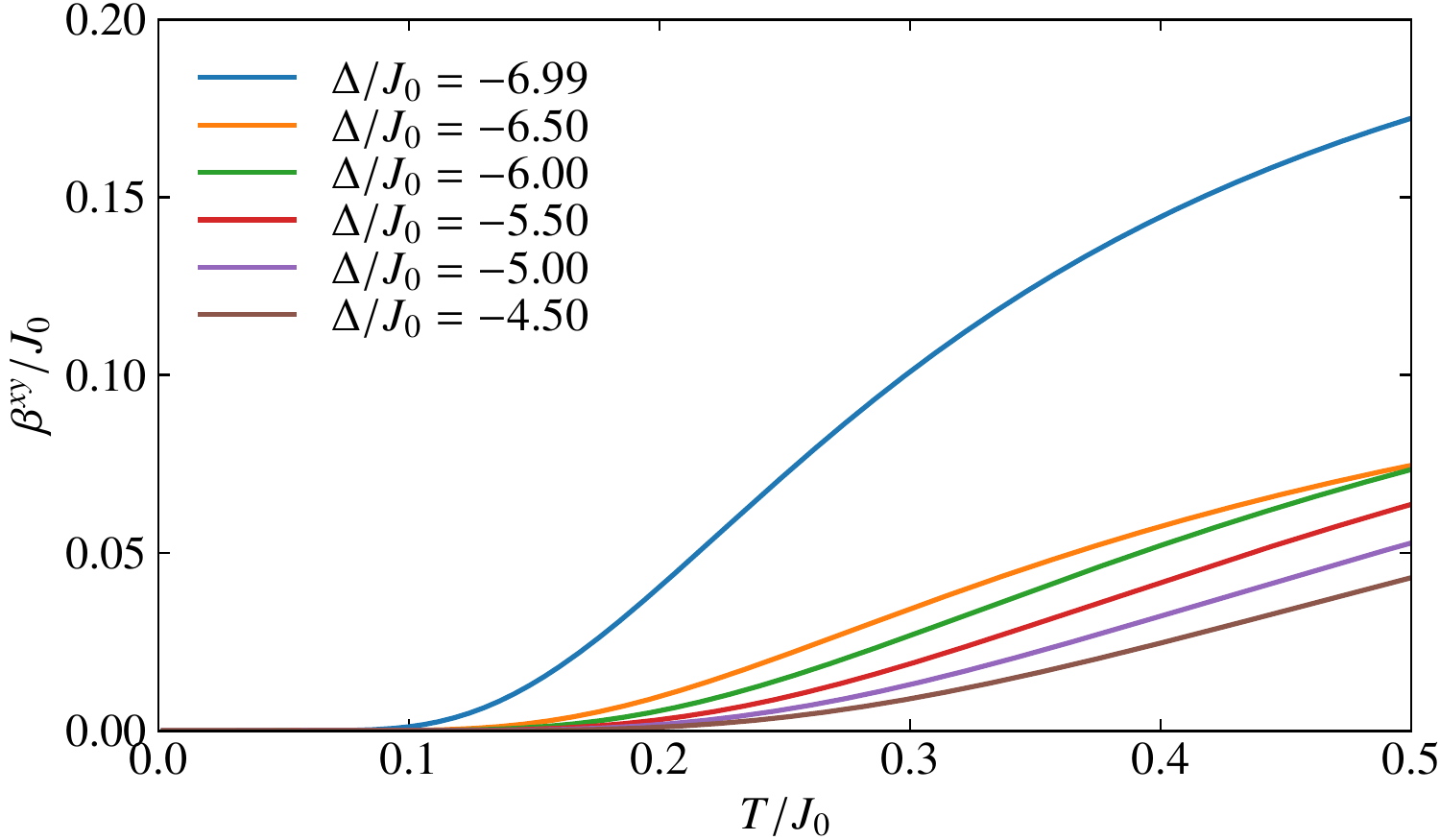}
\caption{
  Temperature dependence of the coefficient of the thermopolarization at $J_1/J_0=0.5$ and $A/J_0=1$.
}
\label{fig:tdep}
\end{center}
\end{figure}

The enhancement around the critical point of $\Delta$ is also observed in the temperature dependence.
As shown in Fig.~\ref{fig:tdep}, $\beta^{xy}$ increases with increasing temperature and takes a large value when $\Delta$ approaches the critical value $-\Delta_c=-7J_0$ for $A/J_0=1$.
In particular, at $\Delta/J_0=-6.99$, $\beta^{xy}$ grows around $T/J_0=0.12$, which is lower than the temperatures in the other cases.
This suggests that the enhancement of $\beta^{xy}$ around the critical region originates from the small gap in the low-energy excitations. 
On the other hand, around the critical point in the positive $\Delta$, the enhancement of $\beta^{xy}$ is not observed even for the small $A$ [$\beta^{xy}$ is almost zero for $\Delta>0$ at $A/J_0=1$ as shown in Fig.~\ref{fig:g_beta}(b)].
The asymmetry is due to the presence of the $p$ orbital degeneracy; the local level of the doubly degenerate $p$ orbitals is lower than that of the $s$ orbital for $\Delta<0$ at $A=0$, but the nondegenerate $s$ orbital is the local ground state for $\Delta>0$.

To examine the impact of the fluctuating $p$ orbitals on enhancing the thermopolarization, we introduce a simple transverse Ising model on a zigzag chain, ${\cal H}_{\rm TI}$ (see the details in Appendix~\ref{app:TI}).
There are two local states at each site in this model, unlike the present Hamiltonian with the three local states.
We find that the coefficient of the off-diagonal thermopolarization is symmetric for the transverse field as well as the order parameter in the transverse Ising model (Fig.~\ref{fig:zigzag_beta}).
This is in stark contrast to the present three-orbital model with small $A$, as shown in the main panel of Fig.~\ref{fig:g_beta}(b).
Moreover, the absolute value of the coefficient in the transverse Ising model is significantly small compared with the energy scale of the interaction, even in the vicinity of the critical points.
Indeed, similar behavior is observed in the three-orbital model with large anisotropy, regarded as a two-orbital model like the transverse Ising model.
The inset of Fig.~\ref{fig:g_beta}(b) shows $\beta^{xy}$ for $A/J_0=100$ at several temperatures.
The symmetric $\Delta$ dependence and the order of the peak value around the critical points appear to be common to those of the transverse Ising model.
These results indicate that the $p$ orbitals play an essential role in enhancing the thermopolarization.

\subsection{Elementary excitation spectrum}
\label{sec:result_excitations}

\begin{figure}[t]
\begin{center}
\includegraphics[width=\columnwidth,clip]{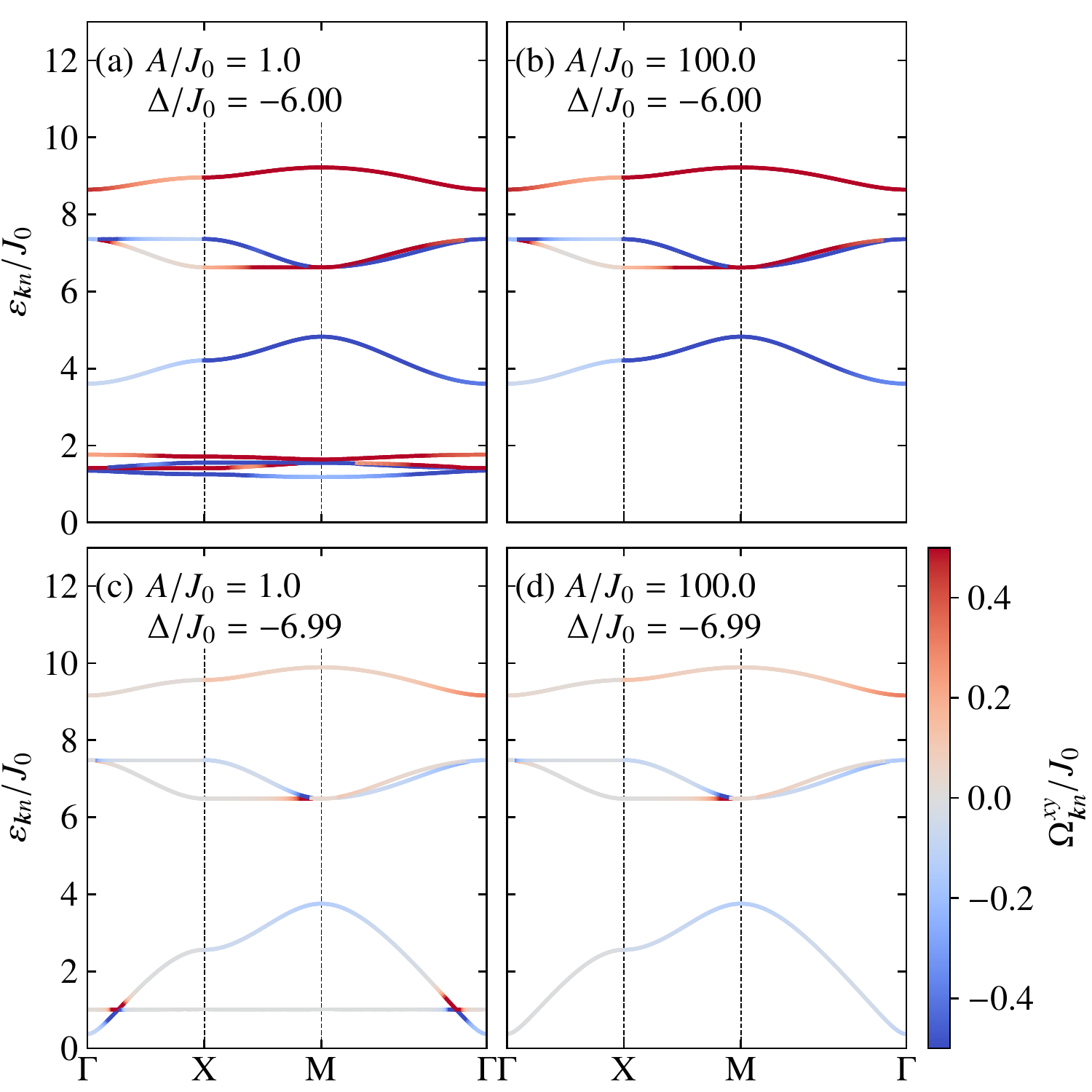}
\caption{
  (a),(b) Dispersion relations of the collective modes from the toroidal-type electric dipole order for (a) $A/J_0=1$ and (b) $A/J_0=100$ at $\Delta/J_0=-6$.
  The color of the lines represents the value of $\Omega_{\bm{k}n}$ for the corresponding excitation.
  (c),(d) Corresponding plots for $\Delta/J_0=-6.99$.
  $J_1/J_0$ is fixed to $0.5$.
  The wave-vector points X and M denote $\bm{k}=(\pi,0)$ and $(\pi,\pi)$, respectively.
}
\label{fig:omega_spec}
\end{center}
\end{figure}

The effect of the fluctuating $p$ orbitals can be clarified by examining the excitation spectrum from the ground state.
Figure~\ref{fig:omega_spec} shows the dispersion relations $\varepsilon_{\bm{k}n}$ of the collective modes and contributions from the corresponding branch to the thermopolarization, $\Omega^{xy}_{\bm{k}n}$.
The dispersion relations for the small and large values of the anisotropy at $\Delta/J_0=-6$ are presented in Figs.~\ref{fig:omega_spec}(a) and \ref{fig:omega_spec}(b), respectively.
At $A/J_0=1$, there are four high-energy branches above $3J_0$ and low-energy excitations below $2J_0$ with small dispersions.
The former are almost unchanged by the large anisotropy, but the latter disappear in the case with $A/J_0=100$, which are located around the higher-energy region scaled by $A$.
These results indicate that the four dispersive branches around $6J_0$ are interpreted as longitudinal modes varying the amplitude of the electric dipoles, which originate from transitions between the $s$ and $p$ orbitals.
This contribution is insensitive to the anisotropy because the local level splitting between the $s$ and $p_{\parallel}$ is independent of $A$.

On the other hand, the anisotropy $A$ lifts the degeneracy of the local $p$ orbitals and yields the energy splitting between the $p_{\parallel}$ and $p_{\perp}$ orbitals.
Since the energy gap of the low-energy modes below $3J_0$ in Fig.~\ref{fig:omega_spec}(a) depends on the anisotropy $A$, these modes are understood as the fluctuation between the two orbitals, corresponding to the transverse modes changing the direction of the electric dipoles.
The low-energy transverse modes are associated with nonzero $\Omega^{xy}_{\bm{k}n}$, which leads to a significant value of $\beta^{xy}$ compared to that in the case with the large local anisotropy.
Moreover, we also find the negative $\Omega^{xy}_{\bm{k}n}$ in the lowest energy branch in Fig.~\ref{fig:omega_spec}(a).
This results in a positive value of $\beta^{xy}$ because of the negative sign in Eq.~\eqref{eq:betaxy}.

Next, we focus on the vicinity of the critical point at $\Delta/J_0=-7$.
Figure~\ref{fig:omega_spec}(c) shows the excitation spectrum at $\Delta/J_0=-6.99$ and $A/J_0=1$.
In this case, there are four almost non-dispersive branches at $\simeq J_0$, which are transverse modes.
This energy corresponds to the value of the anisotropy $A$, and these branches are not observed at $A/J_0=100$ in the energy window of Fig.~\ref{fig:omega_spec}(d).
We find that the averaged value of $\Omega^{xy}_{\bm{k}n}$ for the transverse modes at $A/J_0=1$ is almost zero, and hence, these modes have only a limited effect on $\beta^{xy}$.
However, around the crossing points between the transverse and longitudinal modes, $\Omega^{xy}_{\bm{k}n}$ takes a large value.
The low-energy transverse modes yield this effect as it is not observed in Fig.~\ref{fig:omega_spec}(d).
In particular, $\Omega^{xy}_{\bm{k}n}$ for the low-energy longitudinal mode below $\varepsilon/J_0\sim 1$ takes a considerable negative value, which results in the substantial enhancement of the thermopolarization in the vicinity of the critical point.

\begin{figure}[t]
  \begin{center}
  \includegraphics[width=\columnwidth,clip]{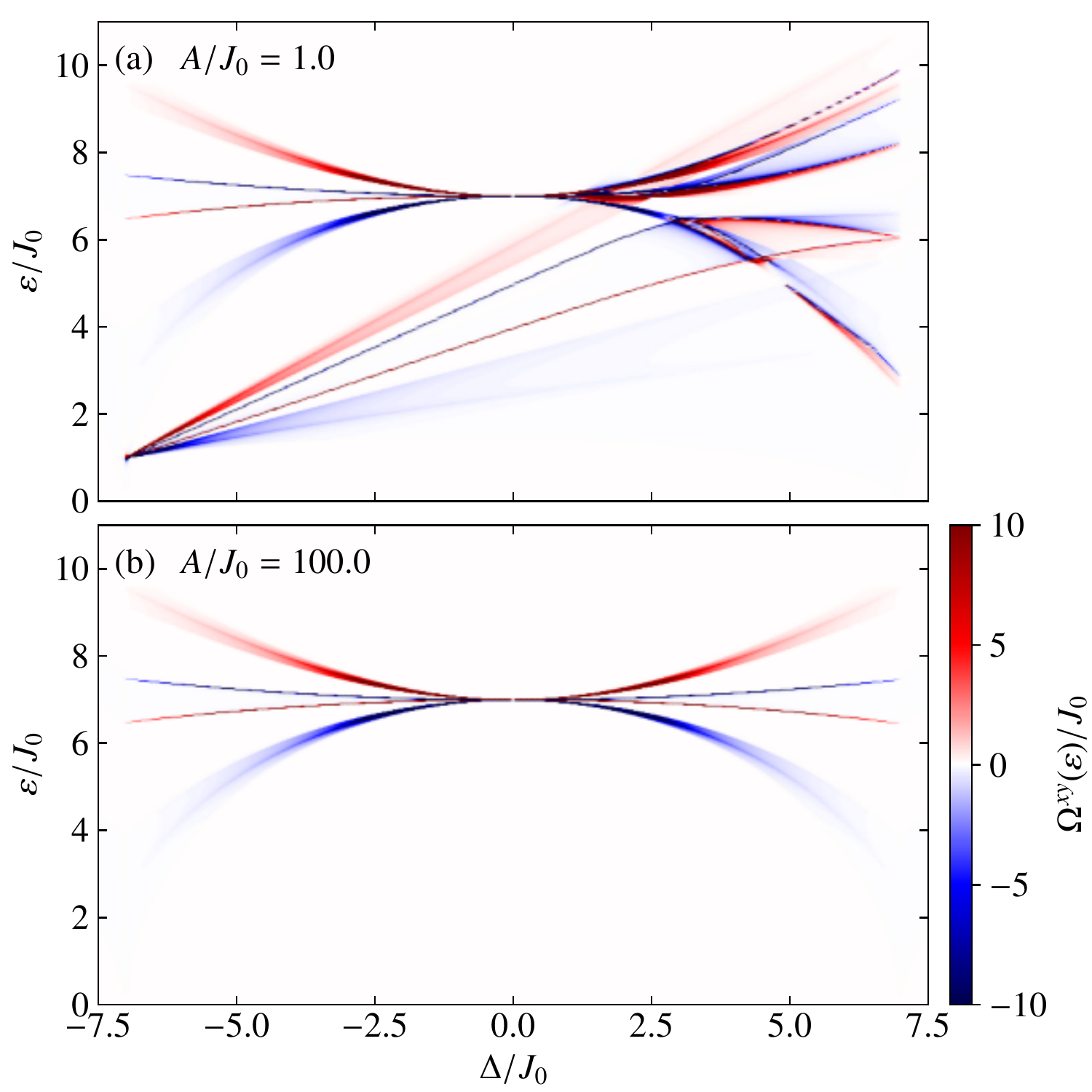}
  \caption{
    Contour map of $\Omega^{xy}(\varepsilon)$ on the plane of $\Delta$ and $\varepsilon$ for $A/J_0=1$ and $A/J_0=100$.
    $J_1/J_0$ is fixed to $0.5$.
  }
  \label{fig:map}
  \end{center}
\end{figure}

To see this effect more clearly, we calculate the $\Delta$ dependence of the density of $\Omega^{xy}_{\bm{k}n}$, which is temperature-independent and defined as
\begin{align}
  \Omega^{xy}(\varepsilon)=\frac{1}{V}
\sum_{\bm{k}}\sum_{n=1}^{2M} \Omega_{\bm{k}n}^{xy}\delta(\varepsilon-\varepsilon_{\bm{k}n}).
\end{align}
Using this spectral representation, the coefficient of the thermopolarization is given as
\begin{align}
  \beta^{xy}=
-\int c_1(n(\varepsilon))\Omega^{xy}(\varepsilon) d\varepsilon.
\end{align}
Figures~\ref{fig:map}(a) and \ref{fig:map}(b) show the $\Delta$ dependence of $\Omega^{xy}(\varepsilon)$ at $A/J_0=1$ and $100$, respectively.
In Fig.~\ref{fig:map}(b), we find four branches around $\varepsilon/J_0 = 7$.
These are spread when $|\Delta|$ is large but merges into a single line at $\Delta=0$.
The high-energy structure originates from the longitudinal modes of the local electric dipoles.
In the case of the small anisotropy ($A/J_0=1$) in Fig~\ref{fig:map}(a), we find the low-energy structure in addition to the high-energy branches.
The energy increases linearly for the negative $\Delta$ region.
Note that the low-energy structure is asymmetric for $\Delta$ while the high-energy one is symmetric.
This is because the former originates from the transverse modes related to the $p$-orbital fluctuation, which is eliminated by the positive $\Delta$, but the latter from the excitation from the $s$ to $p$ orbital.
In both cases, the lowest-energy part of $\Omega^{xy}(\varepsilon)$ is negative, leading to the positive $\beta^{xy}$.
In the case of the small anisotropy, the transverse modes exist at the lower energy, and therefore, the large thermopolarization is observed in the vicinity of the phase boundary.

\section{Discussion}
\label{sec:discussion}

Here, we estimate the magnitude of the off-diagonal thermopolarization in the present mechanism and discuss the emergence of the electric toroidal dipole using the symmetry argument.
First, we estimate the magnitude of the thermopolarization.
We assume that the order of the local electric dipole $\bm{P}_i$ is scaled by $ea$, where $a$ is the length of the primitive translational vectors.
Then, $\beta^{xy}$ should be scaled by $ek_B/aJ_0$.
When $J_0\sim 1$~meV, $\beta^{xy}$ is approximately given as the order of $10^{-10}$~CK$^{-1}$m$^{-1}$.
In this situation, the thermal gradient with $|\nabla T|\sim 1$~K/cm is expected to induce the electric polarization density with the order of $10^{-2}$~$\mu$C/m$^2$.
It might be relatively small to observe the emergent polarization experimentally.
However, since the polarization originating from the disproportionation of a local electronic cloud is often accompanied by the lattice distortion, we expect a more significant value of $\beta^{xy}$ in real materials.
Moreover, the value might also be enhanced by increasing the thermal gradient and considering systems with smaller energy scales.
This effect could be observed in the materials with the ferro-type electric-toroidal dipole order.
The candidate materials are the compounds with the ferroaxial order for lattice distortions, such as  CaMn$_7$O$_{12}$~\cite{Johnson2012}, NiTiO$_{3}$~\cite{hayashida2020visualization}, RbFe(MoO$_{4}$)$_{2}$~\cite{jin2020observation}, and Ca$_{5}$Ir$_{3}$O$_{12}$\cite{Hanate2021}.
It is desired to search other materials exhibiting electric-toroidal dipole orders in the electronic origin, which might be controllable via the degrees of freedom intrinsic to electrons, such as charge and spin.
The candidates are not only transition metal oxides but also organic salts.

Next, we discuss the appearance of the off-diagonal thermopolarization using the symmetry argument.
The present two-dimensional system on the square-octagon lattice belongs to the $\textrm{D}_{4\rm h}$ symmetry.
Under this symmetry, the local $s$ and $p$ orbitals correspond to $\textrm{A}_{1g}$ and $\textrm{E}_{u}$, respectively, at each site.
The local Hamiltonian is represented by a $3\times 3$ Hermitian matrix based on the real wave functions, the $s$, $p_x$, and $p_y$ orbitals, and it is decomposed by eight traceless matrices in addition to the unit matrix with the $\textrm{A}_{1g}$ symmetry.
Note that the five of them are real, and three are pure-imaginary matrices.
The former are time-reversal even, and the latter are time-reversal odd.
The local Hamiltonian is decomposed into the following irreducible representations:
\begin{align}
  (\textrm{A}_{1g}\oplus\textrm{E}_{u})&\otimes (\textrm{A}_{1g}\oplus\textrm{E}_{u})\notag\\
&=2\textrm{A}_{1g}^+\oplus\textrm{A}_{2g}^-\oplus\textrm{B}_{1g}^+\oplus\textrm{B}_{2g}^+\oplus\textrm{E}_{u}^+\oplus\textrm{E}_{u}^-,
\label{eq:local_sym}
\end{align}
where the suffix $+(-)$ denotes a time-reversal even (odd) representation.
Among them, $\textrm{E}_{u}^+$ corresponds to the local electric dipole $\bm{P}_i=(P_i^x,P_i^y)$ at site $i$.

In addition to the local symmetry, we need to consider the symmetry of the four-sublattice structure.
This degree of freedom is written as the irreducible representations as follows:
\begin{align}
  \textrm{A}_{1g}\oplus\textrm{B}_{1g}\oplus\textrm{E}_{u}.
  \label{eq:sub_sym}
\end{align}
These correspond to sublattice modulations $(+,+,+,+)$ for $\textrm{A}_{1g}$, $(+,-,+,-)$ for $\textrm{B}_{1g}$, and $(+,0,-,0)$ and $(0, -,0,+)$ for $\textrm{E}_{u}$ in the labels of the sublattice, $(1,2,3,4)$, presented in Fig.~\ref{fig:lattice}.
Here, we consider the direct product of $\textrm{E}_{u}^+$ in Eq.~\eqref{eq:local_sym} and $\textrm{E}_{u}$ in Eq.~\eqref{eq:sub_sym}, which is decomposed into $\textrm{A}_{1g}^+\oplus\textrm{A}_{2g}^+\oplus \textrm{B}_{1g}^+ \oplus \textrm{B}_{2g}^+$.
In these irreducible representations, $\textrm{A}_{2g}^+$ corresponds to the electric-toroidal dipole~\cite{Hayami2018Classification}.
This is intuitively understood as follows:
$P^y$ appearing with the sublattice modulation $(+,0,-,0)$ and $P^x$ appearing with $(0, -,0,+)$ correspond to the polarization arrangement shown in Fig.~\ref{fig:lattice}.

\begin{figure}[t]
\begin{center}
\includegraphics[width=\columnwidth,clip]{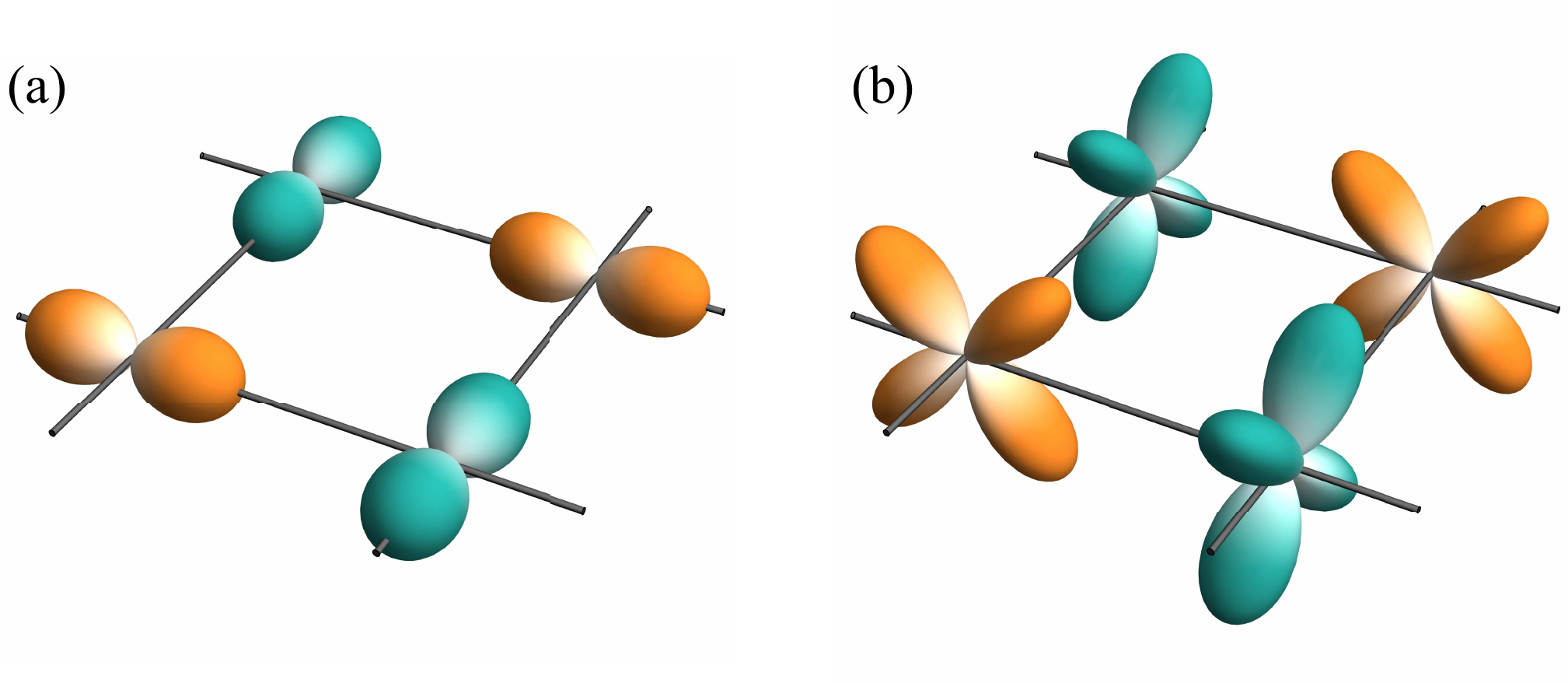}
\caption{
Staggered-type orbital orders composed of (a) $p$ and (b) $d$ orbitals, which are expected to cause the antisymmetric thermopolarization. 
}
\label{fig:orbitals}
\end{center}
\end{figure}

On the other hand, the $\textrm{A}_{2g}^+$ symmetry also appears in the direct product of $\textrm{B}_{2g}^+$ in Eq.~\eqref{eq:local_sym} and $\textrm{B}_{1g}$ in Eq.~\eqref{eq:sub_sym}.
The $\textrm{B}_{2g}^+$ symmetry is derived from the direct product $\textrm{E}_{u}\otimes\textrm{E}_{u}$ within the $p$-orbital sector of the local Hamiltonian, and $\textrm{B}_{1g}$ originates from the sublattice structure.
This suggests that a simple staggered $p$ orbital order on a tetragonal (or square) lattice, which leads to the symmetry lowering from ${\rm D_{4h}}$ to ${\rm C_{4h}}$ in Fig.~\ref{fig:orbitals}(a), also includes the component of an electric toroidal dipole order.
Moreover, under this $p$ orbital order, the antisymmetric off-diagonal thermopolarization should emerge if an $s$ orbital is present near the $p$ orbitals for nonzero matrix elements of the local polarization operator.
Indeed, we could introduce the electric polarization spanning a bond for neighboring sites, which is nonzero, even without the $s$ orbital.
Thus, interacting $p$ models with the staggered orbital order at low-temperature have a potential to exhibit nonzero antisymmetric thermopolarization.
Moreover, a similar argument can be made in $d$ orbital systems with local $\textrm{E}_g$ symmetry [see Fig.~\ref{fig:orbitals}(b)], which will enlarge the range of candidate materials.

\section{Summary}
\label{sec:summary}

In summary, we elucidated that the ferro-type electric-toroidal dipole order induces the antisymmetric thermopolarization by introducing a three-orbital model with $s$ and $p$ orbitals on a two-dimensional lattice.
The mean-field theory suggests that this order emerges when the energy levels of the three local orbitals are close to each other.
By taking account of the fluctuations from the mean fields, we calculate the antisymmetric part of the thermopolarization based on the linear response theory.
This quantity is strongly enhanced around the phase boundary, where the electric-toroidal dipole order disappears, and the $p$-orbital level is lower than that of the $s$ orbital.
The low-energy spectrum clarifies that fluctuations of the $p$ orbitals are crucial for enhancing the thermopolarization.
We also estimated the magnitude of the thermopolarization and discussed the origin based on the symmetry argument.
The present results suggest that the thermal gradient can unveil the electric-toroidal dipole order as a linear response and stimulate further investigations on the electric toroidicity in materials.
On the other hand, our model might be too simple to compare the real compounds directly.
A more realistic model is desired to be proposed, but it is a future issue.

\begin{acknowledgments}
The authors thank H.~Kusunose for fruitful discussions.
Parts of the numerical calculations were performed in the supercomputing
systems in ISSP, the University of Tokyo.
  This work was supported by Grant-in-Aid for Scientific Research from
  JSPS, KAKENHI Grant Nos. JP19K03752, JP19K03742, JP20H00122, JP21H01037, and by JST PRESTO (JPMJPR19L5 and JPMJPR20L8).
\end{acknowledgments}

\appendix

\section{Transverse Ising model on a zigzag chain}
\label{app:TI}

\begin{figure}[t]
\begin{center}
\includegraphics[width=\columnwidth,clip]{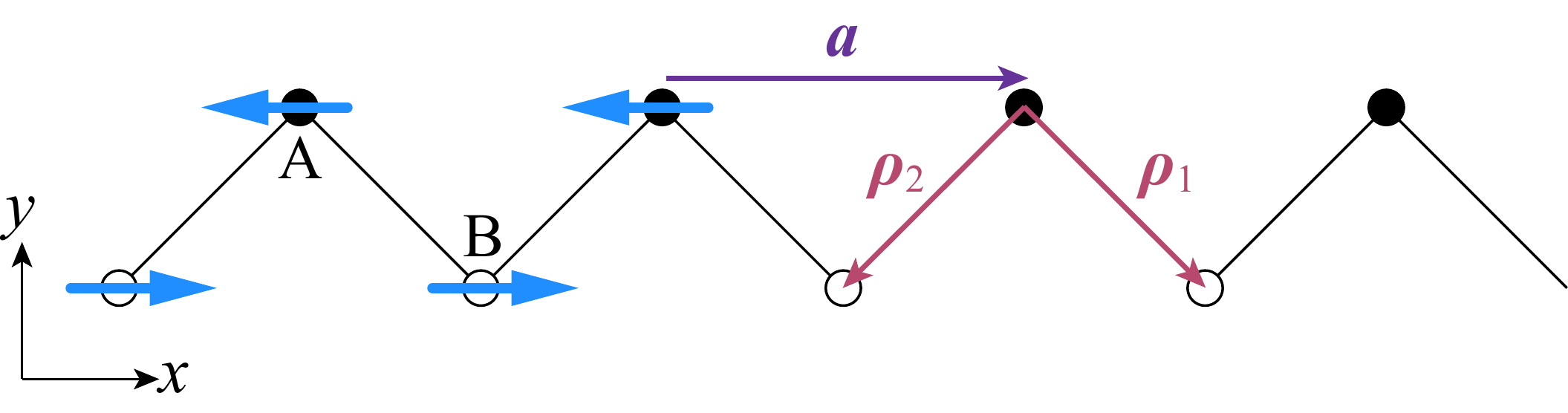}
\caption{
  Schematic picture of the zigzag lattice on which the transverse Ising model is defined.
  The blue arrows represent the electric dipole moments, which exhibit a staggard-type ordering along the $x$ direction.
  $\bm{a}$ is the primitive translational vector, and $\bm{\rho}_1$ and $\bm{\rho}_2$ are the vectors connecting between neighboring sites.
}
\label{fig:zigzag}
\end{center}
\end{figure}

In this appendix, we consider the transverse Ising model on a zigzag lattice as one of the simplest model exhibiting electric toroidal dipole order.
The Hamiltonian is written as
\begin{align}
  {\cal H}_{\rm TI}=J\sum_i \sigma_i^x \sigma_j^x - \Gamma\sum_i \sigma_i^z,~\label{eq:TI}
\end{align}
where $\sigma_i^x$ and $\sigma_i^z$ are the Pauli matrices for two local bases with different parity at site $i$, and the antiferro-type interaction with $J>0$ is assumed.
We regard $\sigma_i^x$ as a local electric-dipole along the $x$ direction, which appears by mixing the two local states.
This means that the two local bases are given by $s$ and $p_x$ orbitals.
The first term means the interactions between electric dipoles, and the second term represents the level splitting, which suppresses the electric dipole moment.

Here, we apply the two-sublattice mean-field approximation to Eq.~\eqref{eq:TI}, where the two-types of local moments are given as
\begin{align}
  -\means{\sigma^x}_{\rm A}=\means{\sigma^x}_{\rm B}&\equiv X \label{eq:appX}\\
  \means{\sigma^z}&\equiv Z,
\end{align}
where we assume the staggard order shown in Fig.~\ref{fig:zigzag} for the A and B sublattices when $X\ne 0$.
In this case, the mean-field energy is given by
\begin{align}
  E_{\rm MF}/N=-JX^2-\Gamma Z,
\end{align}
and the mean-field solution is obtained as
\begin{align}
  \begin{cases}
    Z=\Gamma/\Gamma_c,\quad X=\sqrt{1-Z^2} & \textrm{for}\  |\Gamma|\leq \Gamma_c\\
    Z={\rm sgn}(\Gamma),\quad X=0 &  \textrm{for}\ |\Gamma|>\Gamma_c
  \end{cases},
\end{align}
where $\Gamma_c$ is the critical field given by $\Gamma_c=2J$.

The elementary excitations from the mean-field ground-state are described by bosons as
\begin{align}
  {\cal H}\simeq \tilde{\cal H}
  =E_{\rm MF} +\Delta E\sum_i a_i^\dagger a_i
  -JZ^2\sum_{\means{ij}}\left(a_i^\dagger a_j +a_i a_j + {\rm H.c.}\right),
\end{align}
where $\Delta E=4JX^2+2\Gamma Z$.
By applying the Fourier transformation given as
\begin{align}
a_i=
\begin{cases}
  \sqrt{\frac{2}{N}}\sum_{\bm{k}}a_{\bm{k}}e^{i\bm{k}\cdot\bm{r}_i} & {\rm for}\ i\in A\\
  \sqrt{\frac{2}{N}}\sum_{\bm{k}}b_{\bm{k}}e^{i\bm{k}\cdot\bm{r}_i} & {\rm for}\ i\in B
\end{cases},
\end{align}
the low-energy Hamiltonian is represented as
\begin{align}
\tilde{\cal H}=E_{\rm MF} -\frac{\Delta E N}{2}+\frac{1}{2}\sum_{\bm{k}}
{\cal A}_{\bm{k}}^\dagger H_{\bm{k}} {\cal A}_{\bm{k}},
\end{align}
where $\bm{{\cal A}}_{\bm{k}}=\left(a_{\bm{k}}, b_{\bm{k}}, a_{-\bm{k}}^\dagger, b_{-\bm{k}}^\dagger\right)^T$ and $J_{\bm{k}}=-JZ^2\sum_{\gamma=1,2}e^{i\bm{k}\cdot \bm{\rho}_\gamma}$ with 
$\bm{\rho}_1=(1/2,-1.2)$ and $\bm{\rho}_2=(-1/2,-1.2)$.
The Hamiltonian matrix $H_{\bm{k}}$ is given by
\begin{align}
H_{\bm{k}}=
\begin{pmatrix}
\Delta E & J_{\bm{k}} & 0 & J_{\bm{k}}\\
J_{\bm{k}}^* &\Delta E & J_{\bm{k}}^* & 0\\
0 & J_{\bm{k}} & \Delta E & J_{\bm{k}}\\
J_{\bm{k}}^* & 0 & J_{\bm{k}}^* & \Delta E
\end{pmatrix}.
\end{align}
This matrix is diagonalized by applying the Bogoliubov transformation with paraunitary matrix $T_{\bm{k}}$, and the energies of the two corrective modes are calculated as
\begin{align}
  \varepsilon_{\bm{k}}^{\pm} = \sqrt{\Delta E\left(\Delta E \pm 2|J_{\bm{k}}|\right)}.
\end{align}
The velocity defined in Eq.\eqref{eq:Vy} is represented as
\begin{align}
  {\cal V}_{\bm{k}}^y=T_{\bm{k}}^\dagger\frac{\partial H_{\bm{k}}}{\partial k_y} T_{\bm{k}}
  =
  -\frac{i\Delta E |J_{\bm{k}}|}{2\sqrt{\varepsilon_{\bm{k}}^{+}\varepsilon_{\bm{k}}^{-}}}
  \begin{pmatrix}
    &1&&1\\
    -1&&-1&\\
    &1&&1\\
    -1&&-1&
  \end{pmatrix}.
\end{align}
Moreover, we introduce the the polarization defined by 
\begin{align}
  P^x=\sum_{i}\sigma_i^x.
\end{align}
This is rewritten by using the bosons and approximately given by
\begin{align}
  P^x\simeq 
  2X\sum_{\bm{k}}\left(a_{\bm{k}}^\dagger a_{\bm{k}} - b_{\bm{k}}^\dagger b_{\bm{k}}\right)
  =\frac{1}{2}\sum_{\bm{k}}{\cal A}_{\bm{k}}^\dagger  P_{\bm{k}}^x {\cal A}_{\bm{k}},
\end{align}
where we neglect the linear terms of bosonic operators as it changes the parity of the number of bosons.
$P_{\bm{k}}^x$ is a $4\times 4$ matrix given as
\begin{align}
  P_{\bm{k}}^x=
  \begin{pmatrix}
    2X&&&\\
    &-2X&&\\
    &&2X&\\
    &&&-2X
  \end{pmatrix}.
\end{align}
The matrix ${\cal P}_{\bm{k}}^x$ defined in Eq.~\eqref{eq:Px} is evaluated by ${\cal P}_{\bm{k}}^x=T_{\bm{k}}^\dagger P_{\bm{k}}^x T_{\bm{k}}$.
Using the representations of ${\cal V}_{\bm{k}}^y$ and ${\cal P}_{\bm{k}}^x$, we can calculate $\Omega_{\bm{k}\pm}^{xy}$ in Eq.~\eqref{eq:Omegaxy} as
\begin{align}
  \Omega_{\bm{k}\pm}^{xy}=\pm\frac{X}{2|J_{\bm{k}}|}.
\end{align}
Thus, the coefficient of the transverse thermopolarization, $\beta^{xy}$, is represented as
\begin{align}
  \beta^{xy}
=\frac{1}{V}
\sum_{\bm{k}}
\frac{X}{2|J_{\bm{k}}|}\left\{c_1(n(\varepsilon_{\bm{k}}^-))-c_1(n(\varepsilon_{\bm{k}}^+))\right\}.
\end{align}

\begin{figure}[t]
\begin{center}
\includegraphics[width=\columnwidth,clip]{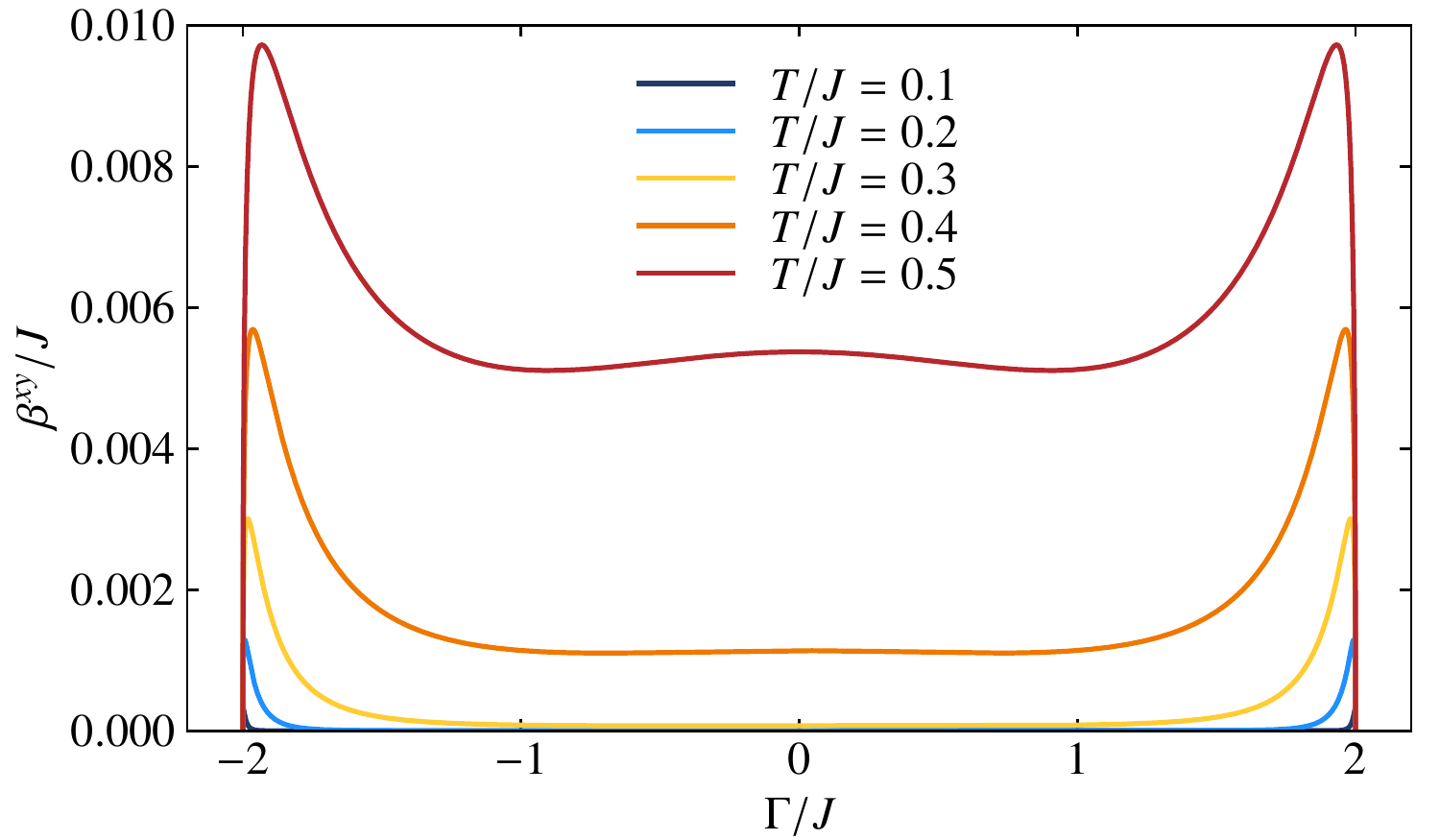}
\caption{
  The transverse field dependence of $\beta^{xy}$ in the transverse Ising model on a zigzag lattice at several temperatures.
  We assume that the volume of the unit cell is unity.
}
\label{fig:zigzag_beta}
\end{center}
\end{figure}

Figure~\ref{fig:zigzag_beta} shows the $\Gamma$ dependence of $\beta^{xy}$.
We find that $\beta^{xy}$ is an even function of $\Gamma$ and increases with increasing temperature.
Moreover, this quantity is enhanced around the critical points $\Gamma/J=\pm 2$, but it takes a small value compared to the energy scale of $J$.
The behavior of $\beta^{xy}$ is distinctly different from that in the three-orbital model on the square-octagon lattice, which is shown in Fig.~\ref{fig:g_beta}(b), while $\means{G}$ as a function of $\Delta$ shown in Fig.~\ref{fig:g_beta}(a) is similar to the $\Gamma$ dependence of the ordered moment $X$ given in Eq.~\eqref{eq:appX}, where $X=\sqrt{1-\Gamma^2/\Gamma_c^2}$ in the ordered phase with $|\Gamma|<\Gamma_c$.

Finally, we discuss the appearance of the electric-toroidal dipole order in the transverse Ising model on the zigzag chain from the viewpoint of the symmetry.
Under the ${\rm D_{2h}}$ symmetry, the localized $s$ and $p_x$ orbitals belongs to $\textrm{A}_g$ and $\textrm{B}_{3u}$.
Then, the local Hamiltonian is given by the $2\times 2$ matrix, which is decomposed into $2\textrm{A}_g^+ \oplus \textrm{B}_{3u}^+ \oplus \textrm{B}_{3u}^-$.
The sublattice degree of freedom is described as $\textrm{A}_g \oplus \textrm{B}_{2u}$.
The $z$ component of the electric-toroidal dipole moment belongs to $\textrm{B}_{1g}^+$, which appears as a part of the direct product of $\textrm{B}_{3u}^+$ in the former and $\textrm{B}_{2u}$ in the latter.

\bibliography{refs}

\begin{thebibliography}{61}%
\makeatletter
\providecommand \@ifxundefined [1]{%
 \@ifx{#1\undefined}
}%
\providecommand \@ifnum [1]{%
 \ifnum #1\expandafter \@firstoftwo
 \else \expandafter \@secondoftwo
 \fi
}%
\providecommand \@ifx [1]{%
 \ifx #1\expandafter \@firstoftwo
 \else \expandafter \@secondoftwo
 \fi
}%
\providecommand \natexlab [1]{#1}%
\providecommand \enquote  [1]{``#1''}%
\providecommand \bibnamefont  [1]{#1}%
\providecommand \bibfnamefont [1]{#1}%
\providecommand \citenamefont [1]{#1}%
\providecommand \href@noop [0]{\@secondoftwo}%
\providecommand \href [0]{\begingroup \@sanitize@url \@href}%
\providecommand \@href[1]{\@@startlink{#1}\@@href}%
\providecommand \@@href[1]{\endgroup#1\@@endlink}%
\providecommand \@sanitize@url [0]{\catcode `\\12\catcode `\$12\catcode
  `\&12\catcode `\#12\catcode `\^12\catcode `\_12\catcode `\%12\relax}%
\providecommand \@@startlink[1]{}%
\providecommand \@@endlink[0]{}%
\providecommand \url  [0]{\begingroup\@sanitize@url \@url }%
\providecommand \@url [1]{\endgroup\@href {#1}{\urlprefix }}%
\providecommand \urlprefix  [0]{URL }%
\providecommand \Eprint [0]{\href }%
\providecommand \doibase [0]{https://doi.org/}%
\providecommand \selectlanguage [0]{\@gobble}%
\providecommand \bibinfo  [0]{\@secondoftwo}%
\providecommand \bibfield  [0]{\@secondoftwo}%
\providecommand \translation [1]{[#1]}%
\providecommand \BibitemOpen [0]{}%
\providecommand \bibitemStop [0]{}%
\providecommand \bibitemNoStop [0]{.\EOS\space}%
\providecommand \EOS [0]{\spacefactor3000\relax}%
\providecommand \BibitemShut  [1]{\csname bibitem#1\endcsname}%
\let\auto@bib@innerbib\@empty
%</preamble>
\bibitem [{\citenamefont {Curie}(1894)}]{curie1894symetrie}%
  \BibitemOpen
  \bibfield  {author} {\bibinfo {author} {\bibfnamefont {P.}~\bibnamefont
  {Curie}},\ }\bibfield  {title} {\bibinfo {title} {Sur la sym{\'e}trie dans
  les ph{\'e}nom{\`e}nes physiques, sym{\'e}trie d'un champ {\'e}lectrique et
  d'un champ magn{\'e}tique},\ }\href
  {https://doi.org/10.1051/jphystap:018940030039300} {\bibfield  {journal}
  {\bibinfo  {journal} {J. Phys. Theor. Appl.}\ }\textbf {\bibinfo {volume}
  {3}},\ \bibinfo {pages} {393} (\bibinfo {year} {1894})}\BibitemShut {NoStop}%
\bibitem [{\citenamefont {Dzyaloshinski\v{i}}(1960)}]{Dzyaloshinski1960}%
  \BibitemOpen
  \bibfield  {author} {\bibinfo {author} {\bibfnamefont {I.~E.}\ \bibnamefont
  {Dzyaloshinski\v{i}}},\ }\bibfield  {title} {\bibinfo {title} {On the
  magneto-electrical effects in antiferromagnets},\ }\href@noop {} {\bibfield
  {journal} {\bibinfo  {journal} {Sov. Phys. JETP}\ }\textbf {\bibinfo {volume}
  {10}},\ \bibinfo {pages} {628} (\bibinfo {year} {1960})}\BibitemShut
  {NoStop}%
\bibitem [{\citenamefont {Astrov}(1960)}]{astrov1960magnetoelectric}%
  \BibitemOpen
  \bibfield  {author} {\bibinfo {author} {\bibfnamefont {D.}~\bibnamefont
  {Astrov}},\ }\bibfield  {title} {\bibinfo {title} {The magnetoelectric effect
  in antiferromagnetics},\ }\href@noop {} {\bibfield  {journal} {\bibinfo
  {journal} {Sov. Phys. JETP}\ }\textbf {\bibinfo {volume} {11}},\ \bibinfo
  {pages} {708} (\bibinfo {year} {1960})}\BibitemShut {NoStop}%
\bibitem [{\citenamefont {Folen}\ \emph {et~al.}(1961)\citenamefont {Folen},
  \citenamefont {Rado},\ and\ \citenamefont {Stalder}}]{Folen1961}%
  \BibitemOpen
  \bibfield  {author} {\bibinfo {author} {\bibfnamefont {V.~J.}\ \bibnamefont
  {Folen}}, \bibinfo {author} {\bibfnamefont {G.~T.}\ \bibnamefont {Rado}},\
  and\ \bibinfo {author} {\bibfnamefont {E.~W.}\ \bibnamefont {Stalder}},\
  }\bibfield  {title} {\bibinfo {title} {Anisotropy of the magnetoelectric
  effect in {{Cr$_2$O$_3$}}},\ }\href
  {https://doi.org/10.1103/PhysRevLett.6.607} {\bibfield  {journal} {\bibinfo
  {journal} {Phys. Rev. Lett.}\ }\textbf {\bibinfo {volume} {6}},\ \bibinfo
  {pages} {607} (\bibinfo {year} {1961})}\BibitemShut {NoStop}%
\bibitem [{\citenamefont {Kimura}\ \emph {et~al.}(2003)\citenamefont {Kimura},
  \citenamefont {Goto}, \citenamefont {Shintani}, \citenamefont {Ishizaka},
  \citenamefont {Arima},\ and\ \citenamefont {Tokura}}]{kimura2003magnetic}%
  \BibitemOpen
  \bibfield  {author} {\bibinfo {author} {\bibfnamefont {T.}~\bibnamefont
  {Kimura}}, \bibinfo {author} {\bibfnamefont {T.}~\bibnamefont {Goto}},
  \bibinfo {author} {\bibfnamefont {H.}~\bibnamefont {Shintani}}, \bibinfo
  {author} {\bibfnamefont {K.}~\bibnamefont {Ishizaka}}, \bibinfo {author}
  {\bibfnamefont {T.-h.}\ \bibnamefont {Arima}},\ and\ \bibinfo {author}
  {\bibfnamefont {Y.}~\bibnamefont {Tokura}},\ }\bibfield  {title} {\bibinfo
  {title} {Magnetic control of ferroelectric polarization},\ }\href
  {https://doi.org/10.1038/nature02018} {\bibfield  {journal} {\bibinfo
  {journal} {nature}\ }\textbf {\bibinfo {volume} {426}},\ \bibinfo {pages}
  {55} (\bibinfo {year} {2003})}\BibitemShut {NoStop}%
\bibitem [{\citenamefont {Fiebig}(2005)}]{fiebig2005revival}%
  \BibitemOpen
  \bibfield  {author} {\bibinfo {author} {\bibfnamefont {M.}~\bibnamefont
  {Fiebig}},\ }\bibfield  {title} {\bibinfo {title} {Revival of the
  magnetoelectric effect},\ }\href {https://doi.org/10.1088/0022-3727/38/8/R01}
  {\bibfield  {journal} {\bibinfo  {journal} {J. Phys. D: Appl. Phys.}\
  }\textbf {\bibinfo {volume} {38}},\ \bibinfo {pages} {R123} (\bibinfo {year}
  {2005})}\BibitemShut {NoStop}%
\bibitem [{\citenamefont {Katsura}\ \emph {et~al.}(2005)\citenamefont
  {Katsura}, \citenamefont {Nagaosa},\ and\ \citenamefont
  {Balatsky}}]{Katsura2005}%
  \BibitemOpen
  \bibfield  {author} {\bibinfo {author} {\bibfnamefont {H.}~\bibnamefont
  {Katsura}}, \bibinfo {author} {\bibfnamefont {N.}~\bibnamefont {Nagaosa}},\
  and\ \bibinfo {author} {\bibfnamefont {A.~V.}\ \bibnamefont {Balatsky}},\
  }\bibfield  {title} {\bibinfo {title} {Spin current and magnetoelectric
  effect in noncollinear magnets},\ }\href
  {https://doi.org/10.1103/PhysRevLett.95.057205} {\bibfield  {journal}
  {\bibinfo  {journal} {Phys. Rev. Lett.}\ }\textbf {\bibinfo {volume} {95}},\
  \bibinfo {pages} {057205} (\bibinfo {year} {2005})}\BibitemShut {NoStop}%
\bibitem [{\citenamefont {Khomskii}(2006)}]{khomskii2006multiferroics}%
  \BibitemOpen
  \bibfield  {author} {\bibinfo {author} {\bibfnamefont {D.~I.}\ \bibnamefont
  {Khomskii}},\ }\bibfield  {title} {\bibinfo {title} {Multiferroics: Different
  ways to combine magnetism and ferroelectricity},\ }\href
  {https://doi.org/10.1016/j.jmmm.2006.01.238} {\bibfield  {journal} {\bibinfo
  {journal} {J. Magn. Magn. Mater.}\ }\textbf {\bibinfo {volume} {306}},\
  \bibinfo {pages} {1} (\bibinfo {year} {2006})}\BibitemShut {NoStop}%
\bibitem [{\citenamefont {Cheong}\ and\ \citenamefont
  {Mostovoy}(2007)}]{cheong2007multiferroics}%
  \BibitemOpen
  \bibfield  {author} {\bibinfo {author} {\bibfnamefont {S.-W.}\ \bibnamefont
  {Cheong}}\ and\ \bibinfo {author} {\bibfnamefont {M.}~\bibnamefont
  {Mostovoy}},\ }\bibfield  {title} {\bibinfo {title} {Multiferroics: a
  magnetic twist for ferroelectricity},\ }\href
  {https://doi.org/10.1038/nmat1804} {\bibfield  {journal} {\bibinfo  {journal}
  {Nat. Mater.}\ }\textbf {\bibinfo {volume} {6}},\ \bibinfo {pages} {13}
  (\bibinfo {year} {2007})}\BibitemShut {NoStop}%
\bibitem [{\citenamefont {Khomskii}(2009)}]{khomskii2009trend}%
  \BibitemOpen
  \bibfield  {author} {\bibinfo {author} {\bibfnamefont {D.}~\bibnamefont
  {Khomskii}},\ }\bibfield  {title} {\bibinfo {title} {Trend: Classifying
  multiferroics: Mechanisms and effects},\ }\href
  {https://doi.org/10.1103/Physics.2.20} {\bibfield  {journal} {\bibinfo
  {journal} {Physics}\ }\textbf {\bibinfo {volume} {2}},\ \bibinfo {pages} {20}
  (\bibinfo {year} {2009})}\BibitemShut {NoStop}%
\bibitem [{\citenamefont {Wang}\ and\ \citenamefont
  {Pang}(2010)}]{wang2010thermally}%
  \BibitemOpen
  \bibfield  {author} {\bibinfo {author} {\bibfnamefont {C.}~\bibnamefont
  {Wang}}\ and\ \bibinfo {author} {\bibfnamefont {M.}~\bibnamefont {Pang}},\
  }\bibfield  {title} {\bibinfo {title} {Thermally induced spin polarization
  and thermal conductivities in a spin--orbit-coupled two-dimensional electron
  gas},\ }\href {https://doi.org/10.1016/j.ssc.2010.06.013} {\bibfield
  {journal} {\bibinfo  {journal} {Solid State Commun.}\ }\textbf {\bibinfo
  {volume} {150}},\ \bibinfo {pages} {1509} (\bibinfo {year}
  {2010})}\BibitemShut {NoStop}%
\bibitem [{\citenamefont {Dyrda\l{}}\ \emph {et~al.}(2013)\citenamefont
  {Dyrda\l{}}, \citenamefont {Inglot}, \citenamefont {Dugaev},\ and\
  \citenamefont {Barna\ifmmode~\acute{s}\else \'{s}\fi{}}}]{Dyrdal2013}%
  \BibitemOpen
  \bibfield  {author} {\bibinfo {author} {\bibfnamefont {A.}~\bibnamefont
  {Dyrda\l{}}}, \bibinfo {author} {\bibfnamefont {M.}~\bibnamefont {Inglot}},
  \bibinfo {author} {\bibfnamefont {V.~K.}\ \bibnamefont {Dugaev}},\ and\
  \bibinfo {author} {\bibfnamefont {J.}~\bibnamefont
  {Barna\ifmmode~\acute{s}\else \'{s}\fi{}}},\ }\bibfield  {title} {\bibinfo
  {title} {Thermally induced spin polarization of a two-dimensional electron
  gas},\ }\href {https://doi.org/10.1103/PhysRevB.87.245309} {\bibfield
  {journal} {\bibinfo  {journal} {Phys. Rev. B}\ }\textbf {\bibinfo {volume}
  {87}},\ \bibinfo {pages} {245309} (\bibinfo {year} {2013})}\BibitemShut
  {NoStop}%
\bibitem [{\citenamefont {Xiao}\ \emph {et~al.}(2016)\citenamefont {Xiao},
  \citenamefont {Li},\ and\ \citenamefont {Ma}}]{xiao2016thermoelectric}%
  \BibitemOpen
  \bibfield  {author} {\bibinfo {author} {\bibfnamefont {C.}~\bibnamefont
  {Xiao}}, \bibinfo {author} {\bibfnamefont {D.}~\bibnamefont {Li}},\ and\
  \bibinfo {author} {\bibfnamefont {Z.}~\bibnamefont {Ma}},\ }\bibfield
  {title} {\bibinfo {title} {Thermoelectric response of spin polarization in
  rashba spintronic systems},\ }\href
  {https://doi.org/10.1007/s11467-016-0566-5} {\bibfield  {journal} {\bibinfo
  {journal} {Frontiers of Physics}\ }\textbf {\bibinfo {volume} {11}},\
  \bibinfo {pages} {117201} (\bibinfo {year} {2016})}\BibitemShut {NoStop}%
\bibitem [{\citenamefont {Dyrda\l{}}\ \emph {et~al.}(2018)\citenamefont
  {Dyrda\l{}}, \citenamefont {Barna\ifmmode~\acute{s}\else \'{s}\fi{}},
  \citenamefont {Dugaev},\ and\ \citenamefont {Berakdar}}]{Dyrdal2018}%
  \BibitemOpen
  \bibfield  {author} {\bibinfo {author} {\bibfnamefont {A.}~\bibnamefont
  {Dyrda\l{}}}, \bibinfo {author} {\bibfnamefont {J.}~\bibnamefont
  {Barna\ifmmode~\acute{s}\else \'{s}\fi{}}}, \bibinfo {author} {\bibfnamefont
  {V.~K.}\ \bibnamefont {Dugaev}},\ and\ \bibinfo {author} {\bibfnamefont
  {J.}~\bibnamefont {Berakdar}},\ }\bibfield  {title} {\bibinfo {title}
  {Thermally induced spin polarization in a magnetized two-dimensional electron
  gas with rashba spin-orbit interaction},\ }\href
  {https://doi.org/10.1103/PhysRevB.98.075307} {\bibfield  {journal} {\bibinfo
  {journal} {Phys. Rev. B}\ }\textbf {\bibinfo {volume} {98}},\ \bibinfo
  {pages} {075307} (\bibinfo {year} {2018})}\BibitemShut {NoStop}%
\bibitem [{\citenamefont {Shitade}\ \emph {et~al.}(2019)\citenamefont
  {Shitade}, \citenamefont {Daido},\ and\ \citenamefont
  {Yanase}}]{Shitade2019quadrupole}%
  \BibitemOpen
  \bibfield  {author} {\bibinfo {author} {\bibfnamefont {A.}~\bibnamefont
  {Shitade}}, \bibinfo {author} {\bibfnamefont {A.}~\bibnamefont {Daido}},\
  and\ \bibinfo {author} {\bibfnamefont {Y.}~\bibnamefont {Yanase}},\
  }\bibfield  {title} {\bibinfo {title} {Theory of spin magnetic quadrupole
  moment and temperature-gradient-induced magnetization},\ }\href
  {https://doi.org/10.1103/PhysRevB.99.024404} {\bibfield  {journal} {\bibinfo
  {journal} {Phys. Rev. B}\ }\textbf {\bibinfo {volume} {99}},\ \bibinfo
  {pages} {024404} (\bibinfo {year} {2019})}\BibitemShut {NoStop}%
\bibitem [{\citenamefont {Bresme}\ \emph {et~al.}(2008)\citenamefont {Bresme},
  \citenamefont {Lervik}, \citenamefont {Bedeaux},\ and\ \citenamefont
  {Kjelstrup}}]{Bresme2008}%
  \BibitemOpen
  \bibfield  {author} {\bibinfo {author} {\bibfnamefont {F.}~\bibnamefont
  {Bresme}}, \bibinfo {author} {\bibfnamefont {A.}~\bibnamefont {Lervik}},
  \bibinfo {author} {\bibfnamefont {D.}~\bibnamefont {Bedeaux}},\ and\ \bibinfo
  {author} {\bibfnamefont {S.}~\bibnamefont {Kjelstrup}},\ }\bibfield  {title}
  {\bibinfo {title} {Water polarization under thermal gradients},\ }\href
  {https://doi.org/10.1103/PhysRevLett.101.020602} {\bibfield  {journal}
  {\bibinfo  {journal} {Phys. Rev. Lett.}\ }\textbf {\bibinfo {volume} {101}},\
  \bibinfo {pages} {020602} (\bibinfo {year} {2008})}\BibitemShut {NoStop}%
\bibitem [{\citenamefont {Wirnsberger}\ \emph {et~al.}(2018)\citenamefont
  {Wirnsberger}, \citenamefont {Dellago}, \citenamefont {Frenkel},\ and\
  \citenamefont {Reinhardt}}]{Wirnsberger2018}%
  \BibitemOpen
  \bibfield  {author} {\bibinfo {author} {\bibfnamefont {P.}~\bibnamefont
  {Wirnsberger}}, \bibinfo {author} {\bibfnamefont {C.}~\bibnamefont
  {Dellago}}, \bibinfo {author} {\bibfnamefont {D.}~\bibnamefont {Frenkel}},\
  and\ \bibinfo {author} {\bibfnamefont {A.}~\bibnamefont {Reinhardt}},\
  }\bibfield  {title} {\bibinfo {title} {Theoretical prediction of thermal
  polarization},\ }\href {https://doi.org/10.1103/PhysRevLett.120.226001}
  {\bibfield  {journal} {\bibinfo  {journal} {Phys. Rev. Lett.}\ }\textbf
  {\bibinfo {volume} {120}},\ \bibinfo {pages} {226001} (\bibinfo {year}
  {2018})}\BibitemShut {NoStop}%
\bibitem [{\citenamefont {Onishi}\ \emph {et~al.}(shed)\citenamefont {Onishi},
  \citenamefont {Isobe},\ and\ \citenamefont {Nagaosa}}]{onishi2021pre}%
  \BibitemOpen
  \bibfield  {author} {\bibinfo {author} {\bibfnamefont {Y.}~\bibnamefont
  {Onishi}}, \bibinfo {author} {\bibfnamefont {H.}~\bibnamefont {Isobe}},\ and\
  \bibinfo {author} {\bibfnamefont {N.}~\bibnamefont {Nagaosa}},\ }\bibfield
  {title} {\bibinfo {title} {Theory of thermoelectric effect in insulators},\
  }\href {http://arxiv.org/abs/2105.08228} {\bibfield  {journal} {\bibinfo
  {journal} {preprint}\ ,\ \bibinfo {pages} {arXiv:2105.08228}} (\bibinfo
  {year} {unpublished})}\BibitemShut {NoStop}%
\bibitem [{\citenamefont {Hayami}\ \emph {et~al.}(2018)\citenamefont {Hayami},
  \citenamefont {Yatsushiro}, \citenamefont {Yanagi},\ and\ \citenamefont
  {Kusunose}}]{Hayami2018Classification}%
  \BibitemOpen
  \bibfield  {author} {\bibinfo {author} {\bibfnamefont {S.}~\bibnamefont
  {Hayami}}, \bibinfo {author} {\bibfnamefont {M.}~\bibnamefont {Yatsushiro}},
  \bibinfo {author} {\bibfnamefont {Y.}~\bibnamefont {Yanagi}},\ and\ \bibinfo
  {author} {\bibfnamefont {H.}~\bibnamefont {Kusunose}},\ }\bibfield  {title}
  {\bibinfo {title} {Classification of atomic-scale multipoles under
  crystallographic point groups and application to linear response tensors},\
  }\href {https://doi.org/10.1103/PhysRevB.98.165110} {\bibfield  {journal}
  {\bibinfo  {journal} {Phys. Rev. B}\ }\textbf {\bibinfo {volume} {98}},\
  \bibinfo {pages} {165110} (\bibinfo {year} {2018})}\BibitemShut {NoStop}%
\bibitem [{\citenamefont {Suzuki}\ \emph {et~al.}(2018)\citenamefont {Suzuki},
  \citenamefont {Ikeda},\ and\ \citenamefont {Oppeneer}}]{Suzuki2018}%
  \BibitemOpen
  \bibfield  {author} {\bibinfo {author} {\bibfnamefont {M.-T.}\ \bibnamefont
  {Suzuki}}, \bibinfo {author} {\bibfnamefont {H.}~\bibnamefont {Ikeda}},\ and\
  \bibinfo {author} {\bibfnamefont {P.~M.}\ \bibnamefont {Oppeneer}},\
  }\bibfield  {title} {\bibinfo {title} {First-principles theory of magnetic
  multipoles in condensed matter systems},\ }\href
  {https://doi.org/10.7566/JPSJ.87.041008} {\bibfield  {journal} {\bibinfo
  {journal} {J. Phys. Soc. Jpn.}\ }\textbf {\bibinfo {volume} {87}},\ \bibinfo
  {pages} {041008} (\bibinfo {year} {2018})}\BibitemShut {NoStop}%
\bibitem [{\citenamefont {Watanabe}\ and\ \citenamefont
  {Yanase}(2018)}]{Watanabe2018group}%
  \BibitemOpen
  \bibfield  {author} {\bibinfo {author} {\bibfnamefont {H.}~\bibnamefont
  {Watanabe}}\ and\ \bibinfo {author} {\bibfnamefont {Y.}~\bibnamefont
  {Yanase}},\ }\bibfield  {title} {\bibinfo {title} {Group-theoretical
  classification of multipole order: Emergent responses and candidate
  materials},\ }\href {https://doi.org/10.1103/PhysRevB.98.245129} {\bibfield
  {journal} {\bibinfo  {journal} {Phys. Rev. B}\ }\textbf {\bibinfo {volume}
  {98}},\ \bibinfo {pages} {245129} (\bibinfo {year} {2018})}\BibitemShut
  {NoStop}%
\bibitem [{\citenamefont {Yatsushiro}\ \emph {et~al.}(2021)\citenamefont
  {Yatsushiro}, \citenamefont {Kusunose},\ and\ \citenamefont
  {Hayami}}]{Yatsushiro2021}%
  \BibitemOpen
  \bibfield  {author} {\bibinfo {author} {\bibfnamefont {M.}~\bibnamefont
  {Yatsushiro}}, \bibinfo {author} {\bibfnamefont {H.}~\bibnamefont
  {Kusunose}},\ and\ \bibinfo {author} {\bibfnamefont {S.}~\bibnamefont
  {Hayami}},\ }\bibfield  {title} {\bibinfo {title} {Multipole classification
  in 122 magnetic point groups for unified understanding of multiferroic
  responses and transport phenomena},\ }\href
  {https://doi.org/10.1103/PhysRevB.104.054412} {\bibfield  {journal} {\bibinfo
   {journal} {Phys. Rev. B}\ }\textbf {\bibinfo {volume} {104}},\ \bibinfo
  {pages} {054412} (\bibinfo {year} {2021})}\BibitemShut {NoStop}%
\bibitem [{\citenamefont {Spaldin}\ \emph {et~al.}(2008)\citenamefont
  {Spaldin}, \citenamefont {Fiebig},\ and\ \citenamefont
  {Mostovoy}}]{spaldin2008toroidal}%
  \BibitemOpen
  \bibfield  {author} {\bibinfo {author} {\bibfnamefont {N.~A.}\ \bibnamefont
  {Spaldin}}, \bibinfo {author} {\bibfnamefont {M.}~\bibnamefont {Fiebig}},\
  and\ \bibinfo {author} {\bibfnamefont {M.}~\bibnamefont {Mostovoy}},\
  }\bibfield  {title} {\bibinfo {title} {The toroidal moment in
  condensed-matter physics and its relation to the magnetoelectric effect},\
  }\href {https://doi.org/10.1088/0953-8984/20/43/434203} {\bibfield  {journal}
  {\bibinfo  {journal} {J. Phys.: Condens. Matter}\ }\textbf {\bibinfo {volume}
  {20}},\ \bibinfo {pages} {434203} (\bibinfo {year} {2008})}\BibitemShut
  {NoStop}%
\bibitem [{\citenamefont {Kopaev}(2009)}]{Kopaev2009}%
  \BibitemOpen
  \bibfield  {author} {\bibinfo {author} {\bibfnamefont {Y.~V.}\ \bibnamefont
  {Kopaev}},\ }\bibfield  {title} {\bibinfo {title} {Toroidal ordering in
  crystals},\ }\href {https://doi.org/10.3367/ufne.0179.200911d.1175}
  {\bibfield  {journal} {\bibinfo  {journal} {Physics-Uspekhi}\ }\textbf
  {\bibinfo {volume} {52}},\ \bibinfo {pages} {1111} (\bibinfo {year}
  {2009})}\BibitemShut {NoStop}%
\bibitem [{\citenamefont {Dubovik}\ and\ \citenamefont
  {Tugushev}(1990)}]{dubovik1990toroid}%
  \BibitemOpen
  \bibfield  {author} {\bibinfo {author} {\bibfnamefont {V.}~\bibnamefont
  {Dubovik}}\ and\ \bibinfo {author} {\bibfnamefont {V.}~\bibnamefont
  {Tugushev}},\ }\bibfield  {title} {\bibinfo {title} {Toroid moments in
  electrodynamics and solid-state physics},\ }\href
  {https://doi.org/10.1016/0370-1573(90)90042-Z} {\bibfield  {journal}
  {\bibinfo  {journal} {Phys. Rep.}\ }\textbf {\bibinfo {volume} {187}},\
  \bibinfo {pages} {145} (\bibinfo {year} {1990})}\BibitemShut {NoStop}%
\bibitem [{\citenamefont {Johnson}\ \emph {et~al.}(2012)\citenamefont
  {Johnson}, \citenamefont {Chapon}, \citenamefont {Khalyavin}, \citenamefont
  {Manuel}, \citenamefont {Radaelli},\ and\ \citenamefont
  {Martin}}]{Johnson2012}%
  \BibitemOpen
  \bibfield  {author} {\bibinfo {author} {\bibfnamefont {R.~D.}\ \bibnamefont
  {Johnson}}, \bibinfo {author} {\bibfnamefont {L.~C.}\ \bibnamefont {Chapon}},
  \bibinfo {author} {\bibfnamefont {D.~D.}\ \bibnamefont {Khalyavin}}, \bibinfo
  {author} {\bibfnamefont {P.}~\bibnamefont {Manuel}}, \bibinfo {author}
  {\bibfnamefont {P.~G.}\ \bibnamefont {Radaelli}},\ and\ \bibinfo {author}
  {\bibfnamefont {C.}~\bibnamefont {Martin}},\ }\bibfield  {title} {\bibinfo
  {title} {Giant improper ferroelectricity in the ferroaxial magnet
  ${\mathrm{camn}}_{7}{\mathbf{o}}_{12}$},\ }\href
  {https://doi.org/10.1103/PhysRevLett.108.067201} {\bibfield  {journal}
  {\bibinfo  {journal} {Phys. Rev. Lett.}\ }\textbf {\bibinfo {volume} {108}},\
  \bibinfo {pages} {067201} (\bibinfo {year} {2012})}\BibitemShut {NoStop}%
\bibitem [{\citenamefont {Hlinka}\ \emph {et~al.}(2016)\citenamefont {Hlinka},
  \citenamefont {Privratska}, \citenamefont {Ondrejkovic},\ and\ \citenamefont
  {Janovec}}]{Hlinka2016}%
  \BibitemOpen
  \bibfield  {author} {\bibinfo {author} {\bibfnamefont {J.}~\bibnamefont
  {Hlinka}}, \bibinfo {author} {\bibfnamefont {J.}~\bibnamefont {Privratska}},
  \bibinfo {author} {\bibfnamefont {P.}~\bibnamefont {Ondrejkovic}},\ and\
  \bibinfo {author} {\bibfnamefont {V.}~\bibnamefont {Janovec}},\ }\bibfield
  {title} {\bibinfo {title} {Symmetry guide to ferroaxial transitions},\ }\href
  {https://doi.org/10.1103/PhysRevLett.116.177602} {\bibfield  {journal}
  {\bibinfo  {journal} {Phys. Rev. Lett.}\ }\textbf {\bibinfo {volume} {116}},\
  \bibinfo {pages} {177602} (\bibinfo {year} {2016})}\BibitemShut {NoStop}%
\bibitem [{\citenamefont {Cheong}\ \emph {et~al.}(2018)\citenamefont {Cheong},
  \citenamefont {Talbayev}, \citenamefont {Kiryukhin},\ and\ \citenamefont
  {Saxena}}]{cheong2018broken}%
  \BibitemOpen
  \bibfield  {author} {\bibinfo {author} {\bibfnamefont {S.-W.}\ \bibnamefont
  {Cheong}}, \bibinfo {author} {\bibfnamefont {D.}~\bibnamefont {Talbayev}},
  \bibinfo {author} {\bibfnamefont {V.}~\bibnamefont {Kiryukhin}},\ and\
  \bibinfo {author} {\bibfnamefont {A.}~\bibnamefont {Saxena}},\ }\bibfield
  {title} {\bibinfo {title} {Broken symmetries, non-reciprocity, and
  multiferroicity},\ }\href {https://doi.org/10.1038/s41535-018-0092-5}
  {\bibfield  {journal} {\bibinfo  {journal} {npj Quantum Materials}\ }\textbf
  {\bibinfo {volume} {3}},\ \bibinfo {pages} {1} (\bibinfo {year}
  {2018})}\BibitemShut {NoStop}%
\bibitem [{\citenamefont {Hayami}\ and\ \citenamefont
  {Kusunose}(2018)}]{Hayami2018Microscopic}%
  \BibitemOpen
  \bibfield  {author} {\bibinfo {author} {\bibfnamefont {S.}~\bibnamefont
  {Hayami}}\ and\ \bibinfo {author} {\bibfnamefont {H.}~\bibnamefont
  {Kusunose}},\ }\bibfield  {title} {\bibinfo {title} {Microscopic description
  of electric and magnetic toroidal multipoles in hybrid orbitals},\ }\href
  {https://doi.org/10.7566/JPSJ.87.033709} {\bibfield  {journal} {\bibinfo
  {journal} {J. Phys. Soc. Jpn.}\ }\textbf {\bibinfo {volume} {87}},\ \bibinfo
  {pages} {033709} (\bibinfo {year} {2018})}\BibitemShut {NoStop}%
\bibitem [{\citenamefont {Hanawa}\ \emph {et~al.}(2001)\citenamefont {Hanawa},
  \citenamefont {Muraoka}, \citenamefont {Tayama}, \citenamefont {Sakakibara},
  \citenamefont {Yamaura},\ and\ \citenamefont {Hiroi}}]{Hanawa2001}%
  \BibitemOpen
  \bibfield  {author} {\bibinfo {author} {\bibfnamefont {M.}~\bibnamefont
  {Hanawa}}, \bibinfo {author} {\bibfnamefont {Y.}~\bibnamefont {Muraoka}},
  \bibinfo {author} {\bibfnamefont {T.}~\bibnamefont {Tayama}}, \bibinfo
  {author} {\bibfnamefont {T.}~\bibnamefont {Sakakibara}}, \bibinfo {author}
  {\bibfnamefont {J.}~\bibnamefont {Yamaura}},\ and\ \bibinfo {author}
  {\bibfnamefont {Z.}~\bibnamefont {Hiroi}},\ }\bibfield  {title} {\bibinfo
  {title} {Superconductivity at 1~{K} in {Cd$_2$Re$_2$O$_7$}},\ }\href
  {https://doi.org/10.1103/PhysRevLett.87.187001} {\bibfield  {journal}
  {\bibinfo  {journal} {Phys. Rev. Lett.}\ }\textbf {\bibinfo {volume} {87}},\
  \bibinfo {pages} {187001} (\bibinfo {year} {2001})}\BibitemShut {NoStop}%
\bibitem [{\citenamefont {Jin}\ \emph {et~al.}(2001)\citenamefont {Jin},
  \citenamefont {He}, \citenamefont {McCall}, \citenamefont {Alexander},
  \citenamefont {Drymiotis},\ and\ \citenamefont {Mandrus}}]{Jin2001}%
  \BibitemOpen
  \bibfield  {author} {\bibinfo {author} {\bibfnamefont {R.}~\bibnamefont
  {Jin}}, \bibinfo {author} {\bibfnamefont {J.}~\bibnamefont {He}}, \bibinfo
  {author} {\bibfnamefont {S.}~\bibnamefont {McCall}}, \bibinfo {author}
  {\bibfnamefont {C.~S.}\ \bibnamefont {Alexander}}, \bibinfo {author}
  {\bibfnamefont {F.}~\bibnamefont {Drymiotis}},\ and\ \bibinfo {author}
  {\bibfnamefont {D.}~\bibnamefont {Mandrus}},\ }\bibfield  {title} {\bibinfo
  {title} {Superconductivity in the correlated pyrochlore
  {Cd$_2$Re$_2$O$_7$}},\ }\href {https://doi.org/10.1103/PhysRevB.64.180503}
  {\bibfield  {journal} {\bibinfo  {journal} {Phys. Rev. B}\ }\textbf {\bibinfo
  {volume} {64}},\ \bibinfo {pages} {180503} (\bibinfo {year}
  {2001})}\BibitemShut {NoStop}%
\bibitem [{\citenamefont {Hiroi}\ \emph {et~al.}(2002)\citenamefont {Hiroi},
  \citenamefont {Yamauchi}, \citenamefont {Yamada}, \citenamefont {Hanawa},
  \citenamefont {Ohishi}, \citenamefont {Shimomura}, \citenamefont {Abliz},
  \citenamefont {Hedo},\ and\ \citenamefont {Uwatoko}}]{Hiroi2002}%
  \BibitemOpen
  \bibfield  {author} {\bibinfo {author} {\bibfnamefont {Z.}~\bibnamefont
  {Hiroi}}, \bibinfo {author} {\bibfnamefont {T.}~\bibnamefont {Yamauchi}},
  \bibinfo {author} {\bibfnamefont {T.}~\bibnamefont {Yamada}}, \bibinfo
  {author} {\bibfnamefont {M.}~\bibnamefont {Hanawa}}, \bibinfo {author}
  {\bibfnamefont {Y.}~\bibnamefont {Ohishi}}, \bibinfo {author} {\bibfnamefont
  {O.}~\bibnamefont {Shimomura}}, \bibinfo {author} {\bibfnamefont
  {M.}~\bibnamefont {Abliz}}, \bibinfo {author} {\bibfnamefont
  {M.}~\bibnamefont {Hedo}},\ and\ \bibinfo {author} {\bibfnamefont
  {Y.}~\bibnamefont {Uwatoko}},\ }\bibfield  {title} {\bibinfo {title}
  {High-pressure study on the superconducting pyrochlore oxide
  {Cd$_2$Re$_2$O$_7$}},\ }\href {https://doi.org/10.1143/JPSJ.71.1553}
  {\bibfield  {journal} {\bibinfo  {journal} {J. Phys. Soc. Jpn.}\ }\textbf
  {\bibinfo {volume} {71}},\ \bibinfo {pages} {1553} (\bibinfo {year}
  {2002})}\BibitemShut {NoStop}%
\bibitem [{\citenamefont {Yamaura}\ and\ \citenamefont
  {Hiroi}(2002)}]{Yamaura2002}%
  \BibitemOpen
  \bibfield  {author} {\bibinfo {author} {\bibfnamefont {J.-I.}\ \bibnamefont
  {Yamaura}}\ and\ \bibinfo {author} {\bibfnamefont {Z.}~\bibnamefont
  {Hiroi}},\ }\bibfield  {title} {\bibinfo {title} {Low temperature symmetry of
  pyrochlore oxide {Cd$_2$Re$_2$O$_7$}},\ }\href
  {https://doi.org/10.1143/JPSJ.71.2598} {\bibfield  {journal} {\bibinfo
  {journal} {J. Phys. Soc. Jpn.}\ }\textbf {\bibinfo {volume} {71}},\ \bibinfo
  {pages} {2598} (\bibinfo {year} {2002})}\BibitemShut {NoStop}%
\bibitem [{\citenamefont {Castellan}\ \emph {et~al.}(2002)\citenamefont
  {Castellan}, \citenamefont {Gaulin}, \citenamefont {van Duijn}, \citenamefont
  {Lewis}, \citenamefont {Lumsden}, \citenamefont {Jin}, \citenamefont {He},
  \citenamefont {Nagler},\ and\ \citenamefont {Mandrus}}]{Castellan2002}%
  \BibitemOpen
  \bibfield  {author} {\bibinfo {author} {\bibfnamefont {J.~P.}\ \bibnamefont
  {Castellan}}, \bibinfo {author} {\bibfnamefont {B.~D.}\ \bibnamefont
  {Gaulin}}, \bibinfo {author} {\bibfnamefont {J.}~\bibnamefont {van Duijn}},
  \bibinfo {author} {\bibfnamefont {M.~J.}\ \bibnamefont {Lewis}}, \bibinfo
  {author} {\bibfnamefont {M.~D.}\ \bibnamefont {Lumsden}}, \bibinfo {author}
  {\bibfnamefont {R.}~\bibnamefont {Jin}}, \bibinfo {author} {\bibfnamefont
  {J.}~\bibnamefont {He}}, \bibinfo {author} {\bibfnamefont {S.~E.}\
  \bibnamefont {Nagler}},\ and\ \bibinfo {author} {\bibfnamefont
  {D.}~\bibnamefont {Mandrus}},\ }\bibfield  {title} {\bibinfo {title}
  {Structural ordering and symmetry breaking in {Cd$_2$Re$_2$O$_7$}},\ }\href
  {https://doi.org/10.1103/PhysRevB.66.134528} {\bibfield  {journal} {\bibinfo
  {journal} {Phys. Rev. B}\ }\textbf {\bibinfo {volume} {66}},\ \bibinfo
  {pages} {134528} (\bibinfo {year} {2002})}\BibitemShut {NoStop}%
\bibitem [{\citenamefont {Kendziora}\ \emph {et~al.}(2005)\citenamefont
  {Kendziora}, \citenamefont {Sergienko}, \citenamefont {Jin}, \citenamefont
  {He}, \citenamefont {Keppens}, \citenamefont {Sales},\ and\ \citenamefont
  {Mandrus}}]{Kendziora2005}%
  \BibitemOpen
  \bibfield  {author} {\bibinfo {author} {\bibfnamefont {C.~A.}\ \bibnamefont
  {Kendziora}}, \bibinfo {author} {\bibfnamefont {I.~A.}\ \bibnamefont
  {Sergienko}}, \bibinfo {author} {\bibfnamefont {R.}~\bibnamefont {Jin}},
  \bibinfo {author} {\bibfnamefont {J.}~\bibnamefont {He}}, \bibinfo {author}
  {\bibfnamefont {V.}~\bibnamefont {Keppens}}, \bibinfo {author} {\bibfnamefont
  {B.~C.}\ \bibnamefont {Sales}},\ and\ \bibinfo {author} {\bibfnamefont
  {D.}~\bibnamefont {Mandrus}},\ }\bibfield  {title} {\bibinfo {title}
  {Goldstone-mode phonon dynamics in the pyrochlore {Cd$_2$Re$_2$O$_7$}},\
  }\href {https://doi.org/10.1103/PhysRevLett.95.125503} {\bibfield  {journal}
  {\bibinfo  {journal} {Phys. Rev. Lett.}\ }\textbf {\bibinfo {volume} {95}},\
  \bibinfo {pages} {125503} (\bibinfo {year} {2005})}\BibitemShut {NoStop}%
\bibitem [{\citenamefont {Bari\ifmmode \check{s}\else
  \v{s}\fi{}i\ifmmode~\acute{c}\else \'{c}\fi{}}\ \emph
  {et~al.}(2003)\citenamefont {Bari\ifmmode \check{s}\else
  \v{s}\fi{}i\ifmmode~\acute{c}\else \'{c}\fi{}}, \citenamefont {Forr\'o},
  \citenamefont {Mandrus}, \citenamefont {Jin}, \citenamefont {He},\ and\
  \citenamefont {Fazekas}}]{Bari2003}%
  \BibitemOpen
  \bibfield  {author} {\bibinfo {author} {\bibfnamefont {N.}~\bibnamefont
  {Bari\ifmmode \check{s}\else \v{s}\fi{}i\ifmmode~\acute{c}\else \'{c}\fi{}}},
  \bibinfo {author} {\bibfnamefont {L.}~\bibnamefont {Forr\'o}}, \bibinfo
  {author} {\bibfnamefont {D.}~\bibnamefont {Mandrus}}, \bibinfo {author}
  {\bibfnamefont {R.}~\bibnamefont {Jin}}, \bibinfo {author} {\bibfnamefont
  {J.}~\bibnamefont {He}},\ and\ \bibinfo {author} {\bibfnamefont
  {P.}~\bibnamefont {Fazekas}},\ }\bibfield  {title} {\bibinfo {title}
  {Electrical properties of {Cd$_2$Re$_2$O$_7$} under pressure},\ }\href
  {https://doi.org/10.1103/PhysRevB.67.245112} {\bibfield  {journal} {\bibinfo
  {journal} {Phys. Rev. B}\ }\textbf {\bibinfo {volume} {67}},\ \bibinfo
  {pages} {245112} (\bibinfo {year} {2003})}\BibitemShut {NoStop}%
\bibitem [{\citenamefont {A.~Sergienko}\ and\ \citenamefont
  {H.~Curnoe}(2003)}]{Sergienko2003}%
  \BibitemOpen
  \bibfield  {author} {\bibinfo {author} {\bibfnamefont {I.}~\bibnamefont
  {A.~Sergienko}}\ and\ \bibinfo {author} {\bibfnamefont {S.}~\bibnamefont
  {H.~Curnoe}},\ }\bibfield  {title} {\bibinfo {title} {Structural order
  parameter in the pyrochlore superconductor {Cd$_2$Re$_2$O$_7$}},\ }\href
  {https://doi.org/10.1143/JPSJ.72.1607} {\bibfield  {journal} {\bibinfo
  {journal} {J. Phys. Soc. Jpn.}\ }\textbf {\bibinfo {volume} {72}},\ \bibinfo
  {pages} {1607} (\bibinfo {year} {2003})}\BibitemShut {NoStop}%
\bibitem [{\citenamefont {C.~Kobayashi}\ \emph {et~al.}(2011)\citenamefont
  {C.~Kobayashi}, \citenamefont {Irie}, \citenamefont {Yamaura}, \citenamefont
  {Hiroi},\ and\ \citenamefont {Murata}}]{Kobayashi2011}%
  \BibitemOpen
  \bibfield  {author} {\bibinfo {author} {\bibfnamefont {T.}~\bibnamefont
  {C.~Kobayashi}}, \bibinfo {author} {\bibfnamefont {Y.}~\bibnamefont {Irie}},
  \bibinfo {author} {\bibfnamefont {J.-i.}\ \bibnamefont {Yamaura}}, \bibinfo
  {author} {\bibfnamefont {Z.}~\bibnamefont {Hiroi}},\ and\ \bibinfo {author}
  {\bibfnamefont {K.}~\bibnamefont {Murata}},\ }\bibfield  {title} {\bibinfo
  {title} {Superconductivity of heavy carriers in the pressure-induced phases
  of {Cd$_2$Re$_2$O$_7$}},\ }\href {https://doi.org/10.1143/JPSJ.80.023715}
  {\bibfield  {journal} {\bibinfo  {journal} {J. Phys. Soc. Jpn.}\ }\textbf
  {\bibinfo {volume} {80}},\ \bibinfo {pages} {023715} (\bibinfo {year}
  {2011})}\BibitemShut {NoStop}%
\bibitem [{\citenamefont {Yamaura}\ \emph {et~al.}(2017)\citenamefont
  {Yamaura}, \citenamefont {Takeda}, \citenamefont {Ikeda}, \citenamefont
  {Hirao}, \citenamefont {Ohishi}, \citenamefont {Kobayashi},\ and\
  \citenamefont {Hiroi}}]{Yamaura2017}%
  \BibitemOpen
  \bibfield  {author} {\bibinfo {author} {\bibfnamefont {J.-i.}\ \bibnamefont
  {Yamaura}}, \bibinfo {author} {\bibfnamefont {K.}~\bibnamefont {Takeda}},
  \bibinfo {author} {\bibfnamefont {Y.}~\bibnamefont {Ikeda}}, \bibinfo
  {author} {\bibfnamefont {N.}~\bibnamefont {Hirao}}, \bibinfo {author}
  {\bibfnamefont {Y.}~\bibnamefont {Ohishi}}, \bibinfo {author} {\bibfnamefont
  {T.~C.}\ \bibnamefont {Kobayashi}},\ and\ \bibinfo {author} {\bibfnamefont
  {Z.}~\bibnamefont {Hiroi}},\ }\bibfield  {title} {\bibinfo {title}
  {Successive spatial symmetry breaking under high pressure in the
  spin-orbit-coupled metal {Cd$_2$Re$_2$O$_7$}},\ }\href
  {https://doi.org/10.1103/PhysRevB.95.020102} {\bibfield  {journal} {\bibinfo
  {journal} {Phys. Rev. B}\ }\textbf {\bibinfo {volume} {95}},\ \bibinfo
  {pages} {020102} (\bibinfo {year} {2017})}\BibitemShut {NoStop}%
\bibitem [{\citenamefont {Hiroi}\ \emph {et~al.}(2018)\citenamefont {Hiroi},
  \citenamefont {Yamaura}, \citenamefont {Kobayashi}, \citenamefont
  {Matsubayashi},\ and\ \citenamefont {Hirai}}]{Hiroi2018}%
  \BibitemOpen
  \bibfield  {author} {\bibinfo {author} {\bibfnamefont {Z.}~\bibnamefont
  {Hiroi}}, \bibinfo {author} {\bibfnamefont {J.-i.}\ \bibnamefont {Yamaura}},
  \bibinfo {author} {\bibfnamefont {T.~C.}\ \bibnamefont {Kobayashi}}, \bibinfo
  {author} {\bibfnamefont {Y.}~\bibnamefont {Matsubayashi}},\ and\ \bibinfo
  {author} {\bibfnamefont {D.}~\bibnamefont {Hirai}},\ }\bibfield  {title}
  {\bibinfo {title} {Pyrochlore oxide superconductor {Cd$_2$Re$_2$O$_7$}
  revisited},\ }\href {https://doi.org/10.7566/JPSJ.87.024702} {\bibfield
  {journal} {\bibinfo  {journal} {J. Phys. Soc. Jpn.}\ }\textbf {\bibinfo
  {volume} {87}},\ \bibinfo {pages} {024702} (\bibinfo {year}
  {2018})}\BibitemShut {NoStop}%
\bibitem [{\citenamefont {Matsubayashi}\ \emph {et~al.}(2018)\citenamefont
  {Matsubayashi}, \citenamefont {Hirai}, \citenamefont {Tokunaga},\ and\
  \citenamefont {Hiroi}}]{Matsubayashi2018}%
  \BibitemOpen
  \bibfield  {author} {\bibinfo {author} {\bibfnamefont {Y.}~\bibnamefont
  {Matsubayashi}}, \bibinfo {author} {\bibfnamefont {D.}~\bibnamefont {Hirai}},
  \bibinfo {author} {\bibfnamefont {M.}~\bibnamefont {Tokunaga}},\ and\
  \bibinfo {author} {\bibfnamefont {Z.}~\bibnamefont {Hiroi}},\ }\bibfield
  {title} {\bibinfo {title} {Formation and control of twin domains in the
  pyrochlore oxide {Cd$_2$Re$_2$O$_7$}},\ }\href
  {https://doi.org/10.7566/JPSJ.87.104604} {\bibfield  {journal} {\bibinfo
  {journal} {J. Phys. Soc. Jpn.}\ }\textbf {\bibinfo {volume} {87}},\ \bibinfo
  {pages} {104604} (\bibinfo {year} {2018})}\BibitemShut {NoStop}%
\bibitem [{\citenamefont {Di~Matteo}\ and\ \citenamefont
  {Norman}(2017)}]{Matteo2017}%
  \BibitemOpen
  \bibfield  {author} {\bibinfo {author} {\bibfnamefont {S.}~\bibnamefont
  {Di~Matteo}}\ and\ \bibinfo {author} {\bibfnamefont {M.~R.}\ \bibnamefont
  {Norman}},\ }\bibfield  {title} {\bibinfo {title} {Nature of the tensor order
  in {Cd$_2$Re$_2$O$_7$}},\ }\href {https://doi.org/10.1103/PhysRevB.96.115156}
  {\bibfield  {journal} {\bibinfo  {journal} {Phys. Rev. B}\ }\textbf {\bibinfo
  {volume} {96}},\ \bibinfo {pages} {115156} (\bibinfo {year}
  {2017})}\BibitemShut {NoStop}%
\bibitem [{\citenamefont {Hayami}\ \emph {et~al.}(2019)\citenamefont {Hayami},
  \citenamefont {Yanagi}, \citenamefont {Kusunose},\ and\ \citenamefont
  {Motome}}]{Hayami2019ElectricToroidal}%
  \BibitemOpen
  \bibfield  {author} {\bibinfo {author} {\bibfnamefont {S.}~\bibnamefont
  {Hayami}}, \bibinfo {author} {\bibfnamefont {Y.}~\bibnamefont {Yanagi}},
  \bibinfo {author} {\bibfnamefont {H.}~\bibnamefont {Kusunose}},\ and\
  \bibinfo {author} {\bibfnamefont {Y.}~\bibnamefont {Motome}},\ }\bibfield
  {title} {\bibinfo {title} {Electric toroidal quadrupoles in the
  spin-orbit-coupled metal
  ${\mathrm{cd}}_{2}{\mathrm{re}}_{2}{\mathrm{o}}_{7}$},\ }\href
  {https://doi.org/10.1103/PhysRevLett.122.147602} {\bibfield  {journal}
  {\bibinfo  {journal} {Phys. Rev. Lett.}\ }\textbf {\bibinfo {volume} {122}},\
  \bibinfo {pages} {147602} (\bibinfo {year} {2019})}\BibitemShut {NoStop}%
\bibitem [{\citenamefont {Hayami}\ \emph {et~al.}(shed)\citenamefont {Hayami},
  \citenamefont {Oiwa},\ and\ \citenamefont {Kusunose}}]{hayami2021pre}%
  \BibitemOpen
  \bibfield  {author} {\bibinfo {author} {\bibfnamefont {S.}~\bibnamefont
  {Hayami}}, \bibinfo {author} {\bibfnamefont {R.}~\bibnamefont {Oiwa}},\ and\
  \bibinfo {author} {\bibfnamefont {H.}~\bibnamefont {Kusunose}},\ }\bibfield
  {title} {\bibinfo {title} {Electric ferro-axial moment as nanometric rotator
  and source of longitudinal spin current},\ }\href
  {http://arxiv.org/abs/2111.10519} {\bibfield  {journal} {\bibinfo  {journal}
  {preprint}\ ,\ \bibinfo {pages} {arXiv:2111.10519}} (\bibinfo {year}
  {unpublished})}\BibitemShut {NoStop}%
\bibitem [{\citenamefont {Hayashida}\ \emph {et~al.}(2020)\citenamefont
  {Hayashida}, \citenamefont {Uemura}, \citenamefont {Kimura}, \citenamefont
  {Matsuoka}, \citenamefont {Morikawa}, \citenamefont {Hirose}, \citenamefont
  {Tsuda}, \citenamefont {Hasegawa},\ and\ \citenamefont
  {Kimura}}]{hayashida2020visualization}%
  \BibitemOpen
  \bibfield  {author} {\bibinfo {author} {\bibfnamefont {T.}~\bibnamefont
  {Hayashida}}, \bibinfo {author} {\bibfnamefont {Y.}~\bibnamefont {Uemura}},
  \bibinfo {author} {\bibfnamefont {K.}~\bibnamefont {Kimura}}, \bibinfo
  {author} {\bibfnamefont {S.}~\bibnamefont {Matsuoka}}, \bibinfo {author}
  {\bibfnamefont {D.}~\bibnamefont {Morikawa}}, \bibinfo {author}
  {\bibfnamefont {S.}~\bibnamefont {Hirose}}, \bibinfo {author} {\bibfnamefont
  {K.}~\bibnamefont {Tsuda}}, \bibinfo {author} {\bibfnamefont
  {T.}~\bibnamefont {Hasegawa}},\ and\ \bibinfo {author} {\bibfnamefont
  {T.}~\bibnamefont {Kimura}},\ }\bibfield  {title} {\bibinfo {title}
  {Visualization of ferroaxial domains in an order-disorder type ferroaxial
  crystal},\ }\href {https://doi.org/10.1038/s41467-020-18408-6} {\bibfield
  {journal} {\bibinfo  {journal} {Nat. commun.}\ }\textbf {\bibinfo {volume}
  {11}},\ \bibinfo {pages} {1} (\bibinfo {year} {2020})}\BibitemShut {NoStop}%
\bibitem [{\citenamefont {Jin}\ \emph {et~al.}(2020)\citenamefont {Jin},
  \citenamefont {Drueke}, \citenamefont {Li}, \citenamefont {Admasu},
  \citenamefont {Owen}, \citenamefont {Day}, \citenamefont {Sun}, \citenamefont
  {Cheong},\ and\ \citenamefont {Zhao}}]{jin2020observation}%
  \BibitemOpen
  \bibfield  {author} {\bibinfo {author} {\bibfnamefont {W.}~\bibnamefont
  {Jin}}, \bibinfo {author} {\bibfnamefont {E.}~\bibnamefont {Drueke}},
  \bibinfo {author} {\bibfnamefont {S.}~\bibnamefont {Li}}, \bibinfo {author}
  {\bibfnamefont {A.}~\bibnamefont {Admasu}}, \bibinfo {author} {\bibfnamefont
  {R.}~\bibnamefont {Owen}}, \bibinfo {author} {\bibfnamefont {M.}~\bibnamefont
  {Day}}, \bibinfo {author} {\bibfnamefont {K.}~\bibnamefont {Sun}}, \bibinfo
  {author} {\bibfnamefont {S.-W.}\ \bibnamefont {Cheong}},\ and\ \bibinfo
  {author} {\bibfnamefont {L.}~\bibnamefont {Zhao}},\ }\bibfield  {title}
  {\bibinfo {title} {Observation of a ferro-rotational order coupled with
  second-order nonlinear optical fields},\ }\href
  {https://doi.org/10.1038/s41567-019-0695-1} {\bibfield  {journal} {\bibinfo
  {journal} {Nat. Phys.}\ }\textbf {\bibinfo {volume} {16}},\ \bibinfo {pages}
  {42} (\bibinfo {year} {2020})}\BibitemShut {NoStop}%
\bibitem [{\citenamefont {{de Gennes}}(1963)}]{DEGENNES1963132}%
  \BibitemOpen
  \bibfield  {author} {\bibinfo {author} {\bibfnamefont {P.}~\bibnamefont {{de
  Gennes}}},\ }\bibfield  {title} {\bibinfo {title} {Collective motions of
  hydrogen bonds},\ }\href
  {https://doi.org/https://doi.org/10.1016/0038-1098(63)90212-6} {\bibfield
  {journal} {\bibinfo  {journal} {Solid State Commun.}\ }\textbf {\bibinfo
  {volume} {1}},\ \bibinfo {pages} {132} (\bibinfo {year} {1963})}\BibitemShut
  {NoStop}%
\bibitem [{\citenamefont {Hemberger}\ \emph {et~al.}(1996)\citenamefont
  {Hemberger}, \citenamefont {Nicklas}, \citenamefont {Viana}, \citenamefont
  {Lunkenheimer}, \citenamefont {Loidl},\ and\ \citenamefont
  {B{\"o}hmer}}]{hemberger1996quantum}%
  \BibitemOpen
  \bibfield  {author} {\bibinfo {author} {\bibfnamefont {J.}~\bibnamefont
  {Hemberger}}, \bibinfo {author} {\bibfnamefont {M.}~\bibnamefont {Nicklas}},
  \bibinfo {author} {\bibfnamefont {R.}~\bibnamefont {Viana}}, \bibinfo
  {author} {\bibfnamefont {P.}~\bibnamefont {Lunkenheimer}}, \bibinfo {author}
  {\bibfnamefont {A.}~\bibnamefont {Loidl}},\ and\ \bibinfo {author}
  {\bibfnamefont {R.}~\bibnamefont {B{\"o}hmer}},\ }\bibfield  {title}
  {\bibinfo {title} {Quantum paraelectric and induced ferroelectric states
  in},\ }\href@noop {} {\bibfield  {journal} {\bibinfo  {journal} {J. Phys.:
  Condens. Matter}\ }\textbf {\bibinfo {volume} {8}},\ \bibinfo {pages} {4673}
  (\bibinfo {year} {1996})}\BibitemShut {NoStop}%
\bibitem [{\citenamefont {Prosandeev}\ \emph {et~al.}(1999)\citenamefont
  {Prosandeev}, \citenamefont {Kleemann}, \citenamefont
  {Westwa\ifmmode~\acute{n}\else \'{n}\fi{}ski},\ and\ \citenamefont
  {Dec}}]{Prosandeev1999}%
  \BibitemOpen
  \bibfield  {author} {\bibinfo {author} {\bibfnamefont {S.~A.}\ \bibnamefont
  {Prosandeev}}, \bibinfo {author} {\bibfnamefont {W.}~\bibnamefont
  {Kleemann}}, \bibinfo {author} {\bibfnamefont {B.}~\bibnamefont
  {Westwa\ifmmode~\acute{n}\else \'{n}\fi{}ski}},\ and\ \bibinfo {author}
  {\bibfnamefont {J.}~\bibnamefont {Dec}},\ }\bibfield  {title} {\bibinfo
  {title} {Quantum paraelectricity in the mean-field approximation},\ }\href
  {https://doi.org/10.1103/PhysRevB.60.14489} {\bibfield  {journal} {\bibinfo
  {journal} {Phys. Rev. B}\ }\textbf {\bibinfo {volume} {60}},\ \bibinfo
  {pages} {14489} (\bibinfo {year} {1999})}\BibitemShut {NoStop}%
\bibitem [{\citenamefont {Onufrieva}(1985)}]{Onufrieva1985}%
  \BibitemOpen
  \bibfield  {author} {\bibinfo {author} {\bibfnamefont {F.~P.}\ \bibnamefont
  {Onufrieva}},\ }\bibfield  {title} {\bibinfo {title} {Low-temperature
  properties of spin systems with tensor order parameters},\ }\href@noop {}
  {\bibfield  {journal} {\bibinfo  {journal} {Zh. Eksp. Teor. Fiz}\ }\textbf
  {\bibinfo {volume} {89}},\ \bibinfo {pages} {2270} (\bibinfo {year}
  {1985})}\BibitemShut {NoStop}%
\bibitem [{\citenamefont {Papanicolaou}(1988)}]{Papanicolaou1988367}%
  \BibitemOpen
  \bibfield  {author} {\bibinfo {author} {\bibfnamefont {N.}~\bibnamefont
  {Papanicolaou}},\ }\bibfield  {title} {\bibinfo {title} {Unusual phases in
  quantum spin-1 systems},\ }\href
  {https://doi.org/http://dx.doi.org/10.1016/0550-3213(88)90073-9} {\bibfield
  {journal} {\bibinfo  {journal} {Nucl. Phys. B}\ }\textbf {\bibinfo {volume}
  {305}},\ \bibinfo {pages} {367 } (\bibinfo {year} {1988})}\BibitemShut
  {NoStop}%
\bibitem [{\citenamefont {Kusunose}\ and\ \citenamefont
  {Kuramoto}(2001)}]{doi:10.1143/JPSJ.70.3076}%
  \BibitemOpen
  \bibfield  {author} {\bibinfo {author} {\bibfnamefont {H.}~\bibnamefont
  {Kusunose}}\ and\ \bibinfo {author} {\bibfnamefont {Y.}~\bibnamefont
  {Kuramoto}},\ }\bibfield  {title} {\bibinfo {title} {Spin-orbital wave
  excitations in orbitally degenerate exchange model with multipolar
  interactions},\ }\href {https://doi.org/10.1143/JPSJ.70.3076} {\bibfield
  {journal} {\bibinfo  {journal} {J. Phys. Soc. Jpn.}\ }\textbf {\bibinfo
  {volume} {70}},\ \bibinfo {pages} {3076} (\bibinfo {year}
  {2001})}\BibitemShut {NoStop}%
\bibitem [{\citenamefont {Shiina}\ \emph {et~al.}(2003)\citenamefont {Shiina},
  \citenamefont {Shiba}, \citenamefont {Thalmeier}, \citenamefont {Takahashi},\
  and\ \citenamefont {Sakai}}]{doi:10.1143/JPSJ.72.1216}%
  \BibitemOpen
  \bibfield  {author} {\bibinfo {author} {\bibfnamefont {R.}~\bibnamefont
  {Shiina}}, \bibinfo {author} {\bibfnamefont {H.}~\bibnamefont {Shiba}},
  \bibinfo {author} {\bibfnamefont {P.}~\bibnamefont {Thalmeier}}, \bibinfo
  {author} {\bibfnamefont {A.}~\bibnamefont {Takahashi}},\ and\ \bibinfo
  {author} {\bibfnamefont {O.}~\bibnamefont {Sakai}},\ }\bibfield  {title}
  {\bibinfo {title} {Dynamics of multipoles and neutron scattering spectra in
  quadrupolar ordering phase of ceb6},\ }\href
  {https://doi.org/10.1143/JPSJ.72.1216} {\bibfield  {journal} {\bibinfo
  {journal} {J. Phys. Soc. Jpn.}\ }\textbf {\bibinfo {volume} {72}},\ \bibinfo
  {pages} {1216} (\bibinfo {year} {2003})}\BibitemShut {NoStop}%
\bibitem [{\citenamefont {Joshi}\ \emph {et~al.}(1999)\citenamefont {Joshi},
  \citenamefont {Ma}, \citenamefont {Mila}, \citenamefont {Shi},\ and\
  \citenamefont {Zhang}}]{PhysRevB.60.6584}%
  \BibitemOpen
  \bibfield  {author} {\bibinfo {author} {\bibfnamefont {A.}~\bibnamefont
  {Joshi}}, \bibinfo {author} {\bibfnamefont {M.}~\bibnamefont {Ma}}, \bibinfo
  {author} {\bibfnamefont {F.}~\bibnamefont {Mila}}, \bibinfo {author}
  {\bibfnamefont {D.~N.}\ \bibnamefont {Shi}},\ and\ \bibinfo {author}
  {\bibfnamefont {F.~C.}\ \bibnamefont {Zhang}},\ }\bibfield  {title} {\bibinfo
  {title} {Elementary excitations in magnetically ordered systems with orbital
  degeneracy},\ }\href {https://doi.org/10.1103/PhysRevB.60.6584} {\bibfield
  {journal} {\bibinfo  {journal} {Phys. Rev. B}\ }\textbf {\bibinfo {volume}
  {60}},\ \bibinfo {pages} {6584} (\bibinfo {year} {1999})}\BibitemShut
  {NoStop}%
\bibitem [{\citenamefont {Murakami}\ \emph {et~al.}(2013)\citenamefont
  {Murakami}, \citenamefont {Oka},\ and\ \citenamefont
  {Aoki}}]{PhysRevB.88.224404}%
  \BibitemOpen
  \bibfield  {author} {\bibinfo {author} {\bibfnamefont {Y.}~\bibnamefont
  {Murakami}}, \bibinfo {author} {\bibfnamefont {T.}~\bibnamefont {Oka}},\ and\
  \bibinfo {author} {\bibfnamefont {H.}~\bibnamefont {Aoki}},\ }\bibfield
  {title} {\bibinfo {title} {Supersolid states in a spin system: Phase diagram
  and collective excitations},\ }\href
  {https://doi.org/10.1103/PhysRevB.88.224404} {\bibfield  {journal} {\bibinfo
  {journal} {Phys. Rev. B}\ }\textbf {\bibinfo {volume} {88}},\ \bibinfo
  {pages} {224404} (\bibinfo {year} {2013})}\BibitemShut {NoStop}%
\bibitem [{\citenamefont {Nasu}\ and\ \citenamefont
  {Ishihara}(2013)}]{PhysRevB.88.205110}%
  \BibitemOpen
  \bibfield  {author} {\bibinfo {author} {\bibfnamefont {J.}~\bibnamefont
  {Nasu}}\ and\ \bibinfo {author} {\bibfnamefont {S.}~\bibnamefont
  {Ishihara}},\ }\bibfield  {title} {\bibinfo {title} {Vibronic excitation
  dynamics in orbitally degenerate correlated electron system},\ }\href
  {https://doi.org/10.1103/PhysRevB.88.205110} {\bibfield  {journal} {\bibinfo
  {journal} {Phys. Rev. B}\ }\textbf {\bibinfo {volume} {88}},\ \bibinfo
  {pages} {205110} (\bibinfo {year} {2013})}\BibitemShut {NoStop}%
\bibitem [{\citenamefont {Colpa}(1978)}]{COLPA1978327}%
  \BibitemOpen
  \bibfield  {author} {\bibinfo {author} {\bibfnamefont {J.~H.~P.}\
  \bibnamefont {Colpa}},\ }\bibfield  {title} {\bibinfo {title}
  {Diagonalization of the quadratic boson hamiltonian},\ }\href
  {https://doi.org/http://dx.doi.org/10.1016/0378-4371(78)90160-7} {\bibfield
  {journal} {\bibinfo  {journal} {Physica A}\ }\textbf {\bibinfo {volume}
  {93}},\ \bibinfo {pages} {327 } (\bibinfo {year} {1978})}\BibitemShut
  {NoStop}%
\bibitem [{\citenamefont {Murakami}\ and\ \citenamefont
  {Okamoto}(2017)}]{Murakami_Okamoto2017}%
  \BibitemOpen
  \bibfield  {author} {\bibinfo {author} {\bibfnamefont {S.}~\bibnamefont
  {Murakami}}\ and\ \bibinfo {author} {\bibfnamefont {A.}~\bibnamefont
  {Okamoto}},\ }\bibfield  {title} {\bibinfo {title} {Thermal hall effect of
  magnons},\ }\href {https://doi.org/10.7566/JPSJ.86.011010} {\bibfield
  {journal} {\bibinfo  {journal} {J. Phys. Soc. Jpn.}\ }\textbf {\bibinfo
  {volume} {86}},\ \bibinfo {pages} {011010} (\bibinfo {year}
  {2017})}\BibitemShut {NoStop}%
\bibitem [{\citenamefont {Shitade}\ and\ \citenamefont
  {Yanase}(2019)}]{Shitade2019}%
  \BibitemOpen
  \bibfield  {author} {\bibinfo {author} {\bibfnamefont {A.}~\bibnamefont
  {Shitade}}\ and\ \bibinfo {author} {\bibfnamefont {Y.}~\bibnamefont
  {Yanase}},\ }\bibfield  {title} {\bibinfo {title} {Magnon
  gravitomagnetoelectric effect in noncentrosymmetric antiferromagnetic
  insulators},\ }\href {https://doi.org/10.1103/PhysRevB.100.224416} {\bibfield
   {journal} {\bibinfo  {journal} {Phys. Rev. B}\ }\textbf {\bibinfo {volume}
  {100}},\ \bibinfo {pages} {224416} (\bibinfo {year} {2019})}\BibitemShut
  {NoStop}%
\bibitem [{\citenamefont {Li}\ \emph {et~al.}(2020)\citenamefont {Li},
  \citenamefont {Mook}, \citenamefont {Raeliarijaona},\ and\ \citenamefont
  {Kovalev}}]{Li_Mook2020}%
  \BibitemOpen
  \bibfield  {author} {\bibinfo {author} {\bibfnamefont {B.}~\bibnamefont
  {Li}}, \bibinfo {author} {\bibfnamefont {A.}~\bibnamefont {Mook}}, \bibinfo
  {author} {\bibfnamefont {A.}~\bibnamefont {Raeliarijaona}},\ and\ \bibinfo
  {author} {\bibfnamefont {A.~A.}\ \bibnamefont {Kovalev}},\ }\bibfield
  {title} {\bibinfo {title} {Magnonic analog of the edelstein effect in
  antiferromagnetic insulators},\ }\href
  {https://doi.org/10.1103/PhysRevB.101.024427} {\bibfield  {journal} {\bibinfo
   {journal} {Phys. Rev. B}\ }\textbf {\bibinfo {volume} {101}},\ \bibinfo
  {pages} {024427} (\bibinfo {year} {2020})}\BibitemShut {NoStop}%
\bibitem [{\citenamefont {Hanate}\ \emph {et~al.}(2021)\citenamefont {Hanate},
  \citenamefont {Hasegawa}, \citenamefont {Hayami}, \citenamefont {Tsutsui},
  \citenamefont {Kawano},\ and\ \citenamefont {Matsuhira}}]{Hanate2021}%
  \BibitemOpen
  \bibfield  {author} {\bibinfo {author} {\bibfnamefont {H.}~\bibnamefont
  {Hanate}}, \bibinfo {author} {\bibfnamefont {T.}~\bibnamefont {Hasegawa}},
  \bibinfo {author} {\bibfnamefont {S.}~\bibnamefont {Hayami}}, \bibinfo
  {author} {\bibfnamefont {S.}~\bibnamefont {Tsutsui}}, \bibinfo {author}
  {\bibfnamefont {S.}~\bibnamefont {Kawano}},\ and\ \bibinfo {author}
  {\bibfnamefont {K.}~\bibnamefont {Matsuhira}},\ }\bibfield  {title} {\bibinfo
  {title} {First observation of superlattice reflections in the hidden order at
  105~{K} of spin-orbit coupled iridium oxide {Ca$_5$Ir$_3$O$_{12}$}},\ }\href
  {https://doi.org/10.7566/JPSJ.90.063702} {\bibfield  {journal} {\bibinfo
  {journal} {J. Phys. Soc. Jpn.}\ }\textbf {\bibinfo {volume} {90}},\ \bibinfo
  {pages} {063702} (\bibinfo {year} {2021})}\BibitemShut {NoStop}%
\end{thebibliography}%

\end{document}